\documentclass[aps,prb,superscriptaddress,twocolumn,longbibliography,floatfix,10pt]{revtex4-2}

\usepackage[normalem]{ulem}
\usepackage{amsfonts,amssymb,dsfont}
\usepackage[sumlimits,intlimits]{amsmath}
\usepackage{graphics}
\usepackage{graphicx}
\usepackage[dvipsnames]{xcolor}
\usepackage{color}
\usepackage{mathrsfs}
\usepackage{textcomp}
\usepackage{verbatim}
\usepackage{bm}
\usepackage{soul}
\usepackage{braket}
\usepackage{times}
\usepackage[T1]{fontenc} 
\usepackage{booktabs}
\usepackage{lipsum}
\usepackage{tabularx}
\DeclareMathOperator{\Tr}{{\rm Tr}}

\usepackage{xr-hyper}
\usepackage[colorlinks=true,citecolor=blue]{hyperref}
\usepackage[capitalise,compress]{cleveref}
\crefrangelabelformat{equation}{\textup{(#3#1#4)}--\textup{(#5#2#6)}}
\crefname{section}{Sec.}{Secs.}
\Crefname{section}{Section}{Sections}

\begin{document}

	\title{Quantum tensor network algorithms for evaluation of spectral functions on quantum computers}

	\author{Michael L.~Wall}
	\thanks{MLW and AR contributed equally}
	\affiliation{Johns Hopkins University Applied Physics Laboratory, Laurel, Maryland 20723, USA}
	\author{Aidan Reilly}
	\thanks{MLW and AR contributed equally}
	\affiliation{Johns Hopkins University Applied Physics Laboratory, Laurel, Maryland 20723, USA}
	\affiliation{Department of Physics, Stanford University, Stanford, CA 94305, USA
	}
	\author{John S. Van Dyke}
	\affiliation{Johns Hopkins University Applied Physics Laboratory, Laurel, Maryland 20723, USA}
	\author{Collin Broholm}
	\affiliation{William H. Miller III Department of Physics \& Astronomy, Johns Hopkins University, Baltimore, Maryland 21218, USA}
	\author{Paraj Titum}
	\affiliation{Johns Hopkins University Applied Physics Laboratory, Laurel, Maryland 20723, USA}
	\affiliation{William H. Miller III Department of Physics \& Astronomy, Johns Hopkins University, Baltimore, Maryland 21218, USA}
	\thanks{Corresponding Author}
	\email{Paraj.Titum@jhuapl.edu}

	%\date{\today}
	
	\begin{abstract}
		We investigate quantum algorithms derived from tensor networks to simulate the  static and dynamic properties of quantum many-body systems.  Using a sequentially prepared quantum circuit representation of a matrix product state (MPS) that we call a quantum tensor network (QTN), we demonstrate algorithms to prepare ground and excited states on a quantum computer and apply them to molecular nanomagnets (MNMs) as a paradigmatic example. In this setting, we develop two approaches for extracting the spectral correlation functions measured in neutron scattering experiments: (a) a generalization of the SWAP test for computing wavefunction overlaps and, (b) a generalization of the notion of matrix product operators (MPOs) to the QTN setting which generates a linear combination of unitaries.  The latter method is discussed in detail for translationally invariant spin-half systems, where it is shown to reduce the qubit resource requirements compared with the SWAP method, and may be generalized to other systems.  We demonstrate the versatility of our approaches by simulating spin-1/2 and spin-3/2 MNMs, with the latter being an experimentally relevant model of a Cr$^{3+}_8$ ring.  Our approach has qubit requirements that are independent of the number of constituents of the many-body system and scale only logarithmically with the bond dimension of the MPS representation, making them appealing for implementation on near-term quantum hardware with mid-circuit measurement and reset.
	\end{abstract}

	\maketitle
	\section{Introduction}\label{sec:intro}
	Tensor networks (TNs) are one of the leading variational approaches for simulating many-body quantum states using classical computing~\cite{Schollwock2011,orus2014practical,chan2016matrix,orus2019tensor,Dukelsky_1998}. TNs provide an efficient way to decompose the high-dimensional representations of these states in a fixed basis into a network of lower-rank tensors, where the geometry and structure of the network dictates the entanglement and correlations possible in the variational ansatzes~\cite{PhysRevLett.100.070502,eisert2013entanglement}.  A variety of TN ansatzes have been developed for different types of many-body quantum states.  Matrix Product States (MPS) are the most widely used ansatz to simulate short-range entangled many-body quantum states in one-dimension ~\cite{white1992density,Schollwock2011}, and other examples include Projected Entangled Pair States (PEPS) for higher-dimensional states~\cite{PhysRevA.70.060302}, tree tensor networks~\cite{verstraete2004renormalization,PhysRevA.74.022320}, and the Multi-Scale Renormalization Ansatz (MERA) for representing quantum critical states~\cite{PhysRevLett.101.110501}. Several classical tensor network~(CTN) algorithms have been developed to compute dynamic, spectral, and out of equilibrium properties of quantum many-body states~\cite{PhysRevB.94.165116,PhysRevLett.91.147902,wall2012out,PhysRevB.91.165112,PhysRevB.83.195115}. More recently, the TN approach has been applied broadly in the field of machine learning, investigating various TN architectures~\cite{cichocki2014tensor,liu2019machine,evenbly2019number,blagoveschensky2020deep,wall2021tree} for various applications including supervised learning~\cite{stoudenmire2016supervised,stoudenmire2018learning,glasser2020probabilistic,reyes2021multi}, data classification~\cite{klus2019tensor,trenti2020quantum,selvan2020tensor,wang2020anomaly,efthymiou2019tensornetwork}, unsupervised learning~\cite{torlai2020quantum}, generative modeling ~\cite{PhysRevB.99.155131,bradley2020modeling,PhysRevX.8.031012} and others~\cite{gillman2020tensor,miller2020tensor,PhysRevE.98.042114}.

	The success of CTN algorithms has motivated the development of quantum algorithms which exploit a mapping of a CTN to a quantum state preparation routine on a quantum computer~\cite{schon2007sequential}.  We refer to these quantum circuit representations of TNs as \emph{quantum tensor networks} (QTNs). QTN algorithms have also been developed for a wide set of applications in quantum many-body physics and machine learning~\cite{biamonte2018quantum,mugel2020dynamic,wall2021tree,Haghshenas2022,dilip2022data,Dborin_2022,Lubasch2020,Lubasch2023}, including some demonstrations on quantum hardware~\cite{grant2018hierarchical,bhatia2019matrix,wall2020Generative,WallPRA2022,wright2022deterministic}. The expressivity of a CTN ansatz is limited by the so-called \emph{bond dimension} $\chi$, which is the largest dimension of any contracted index in the network of low rank tensors defining the ansatz.  However, CTNs that are equivalent to MPSs lend themselves to a QTN representation with \emph{quantum} resource requirements that scale only logarithmically with $\chi$; that is, preparation of a state corresponding to a MPS with bond dimension $\chi$ requires only $\mathcal{O}\left(\log_2\chi\right)$ qubits~\cite{huggins2019towards}.  Furthermore, MPSs can be prepared using a sequential state preparation quantum algorithm~\cite{schon2007sequential} such that the quantum resource requirements are independent of the number of tensors in the ansatz, e.g.~the number of  sites in a 1D lattice model.  The sequential preparation algorithm requires quantum hardware with mid-circuit measurement and reuse (MCMR), a feature that is currently available on a variety of quantum computing platforms~\cite{pino2020demonstration,gaebler2021suppression}.  The exponential growth in the maximum attainable bond dimension with quantum resources has led to a flurry of research activity into QTN algorithms for representing quantum states~\cite{PhysRevResearch.3.033002,yirka2021qubit,foss2021entanglement,chertkov2021holographic,barratt2021parallel,lin2021real,Haghshenas2022}.
	
		\begin{figure*}
		\centering
		\includegraphics[width=\linewidth]{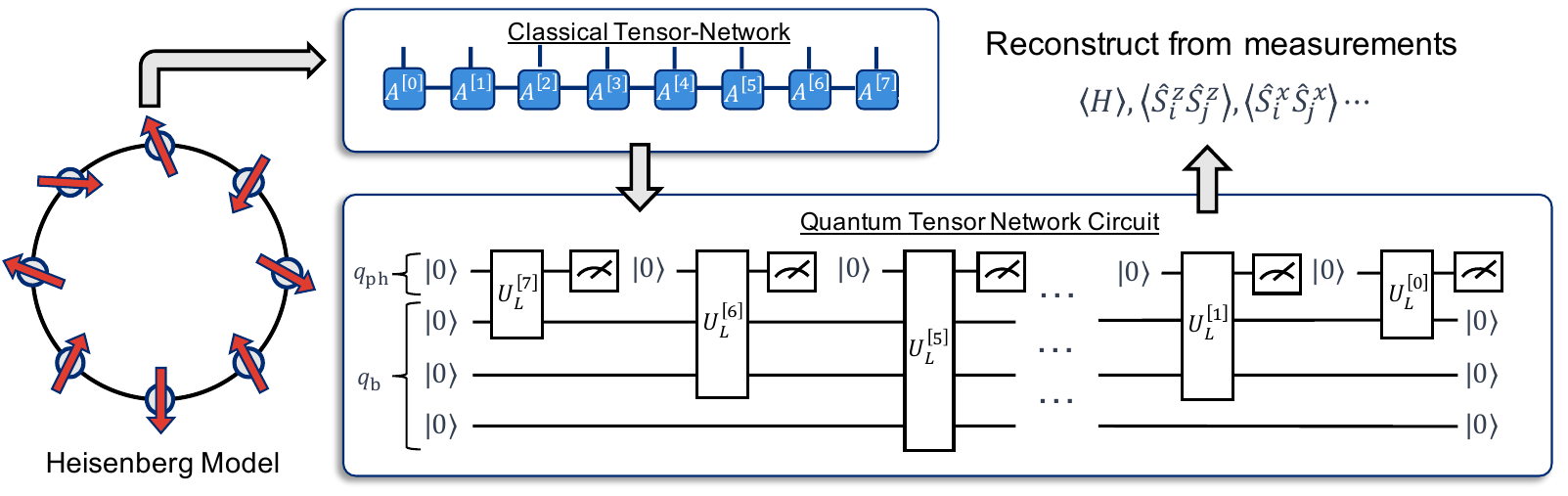}
		\caption{\emph{Schematic steps for preparing ground and excited states of spin models using the quantum tensor network formalism and measuring static correlations.} The ground and excited states of a presumed model description are first obtained as a classical MPS using standard variational techniques.  This classical MPS is mapped through variational compiling to a quantum tensor network. The quantum tensor network circuit is comprised of physical qubits ($q_{\rm ph}$) and bond qubits ($q_{\rm b}$) with the number of physical (bond) qubits corresponding to the base-2 logarithm of the local Hilbert space dimension (bond dimension) of the MPS. The sequential preparation scheme continually measures and resets the physical qubits, and the \emph{static} properties of the many-body quantum states are reconstructed from these measurement records. The schematic prepares a quantum state from an MPS representation in the \emph{left-canonical} form. }\label{fig:schematic}
	\end{figure*}

In this paper, we focus on utilizing QTN algorithms to extract the static and \emph{dynamic} correlation functions of quantum many-body models, such as those underlying quantum magnetism. We focus on the ground and low-lying excited states of spin-$S$ models with Heisenberg exchange coupling in the presence of magnetic fields, a common class of models describing the physics of molecular nanomagnets (MNMs)~\cite{Baker2012,Coronado2019}. MNMs are finite size spin systems and their excitation spectrum is of fundamental and applied interest in condensed matter physics~\cite{Carretta2003,Almeida2009}, with applications in spintronics~\cite{Sanvito2011} and quantum information~\cite{Leuenberger2001,Coronado2019}.
 While small MNMs can be simulated exactly using classical algorithms such as numerical exact-diagonalization, the Hilbert space for representation of the quantum state grows as $\sim (2S+1)^L$ with the number of spins $L$. This exponential growth in Hilbert space makes exact simulation challenging and, therefore, these systems are ideal candidates for simulation with near-term quantum hardware. Of particular interest is simulation of their low-energy excited states, as these can be examined experimentally using spectroscopic methods such as neutron scattering. In fact, the dynamic correlation functions calculated with the QTN algorithms developed in this paper model the neutron-scattering cross-section of the MNM~\cite{Boothroyd2020}.  We provide specific results for the magnetic response function for a spin-3/2 Heisenberg model which is directly comparable to the scattering cross-section obtained from inelastic neutron scattering experiments on the Cr$^{3+}_8$ molecule~\cite{Baker2012}.
	
A schematic description of the steps for computing static correlations with QTNs is given in \cref{fig:schematic}.  Here, CTN representations of the ground and low-lying excited states of a lattice model are obtained using standard variational techniques.  These CTNs are then compiled into QTNs using a greedy variational compilation procedure.  With near-term implementation in mind, we use these variational compilation techniques to identify short-circuit depth representations of the relevant ground and excited states as QTNs, with the accuracy of measured properties in the QTN framework balanced against their circuit depth by the choice of the error tolerance in the compilation procedure.  Once these QTNs are in hand, we extract static and dynamic properties using algorithms that maintain the sequential preparation property of QTNs that makes them particularly quantum resource efficient.  In what follows, we provide two approaches to calculation of the spectral functions. The first approach directly measures a linear combination of many-body operators by using ancilla qubits and controlled operations through a process that can be understood as the QTN analog of matrix product operators (MPOs).  This technique is applied to evaluate momentum space spin operator expectations in a translationally invariant spin-1/2 model, which are then related to the dynamic correlation function through an inverse Fourier transform. The second approach is an extension of the SWAP test algorithm~\cite{Buhrman2001_swaptest} to evaluate the spectral functions.  This approach is more straightforward to adapt to higher-spin representations, but also modestly more quantum resource intensive.  In order to demonstrate the generality of our approaches, we provide example calculations for both spin-1/2 and spin-3/2 models. Since our focus is on developing quantum algorithms to efficiently extract static and dynamic correlations, we do not focus on devising efficient ways to obtain the QTN representations of the molecular states. Thus, we utilize CTN representations of ground states as a starting point for obtaining QTN representations due to the well-developed technology for optimizing CTNs. Future applications of QTNs will likely employ a CTN as a preconditioner for a QTN model that can be scaled to larger effective bond dimension with the addition of qubits and then optimized directly on a quantum device. We note that the algorithms for extracting spectral functions continue to be applicable to this latter approach of generating QTNs.
	
	 This paper is organized as follows. In \cref{sec:model} we introduce the general model for magnetism typically used to describe MNMs as well as the definitions of the  magnetic response function and the magnetic neutron scattering cross-section. In \cref{sec:CTNmethods} we discuss the CTN methods employed in this paper to simulate the properties of the magnetic systems, and generating ground truth data to which we compare the quantum algorithms. \cref{sec:QTN} discusses QTNs, particularly the procedures for preparing and compilation of a CTN state in to a quantum circuit (\cref{subsec:statepre}), and measurement of observables (\cref{subsec:meas}). In \cref{sec:results} we discuss our results obtained from QTN-based simulation of the quantum circuits for the spin models and compare them with the CTN based approaches. In \cref{subsec:spin 1/2} we compare individual eigenstates obtained for the spin-1/2 Heisenberg model and the magnetic response functions. In \cref{subsec:extensions}, we discuss extensions to higher spin models, and discuss results for the experimentally relevant spin-3/2 model describing Cr$^{3+}_8$ molecular nanomagnets.  In \cref{sec:discussion} we provide a summary and outlook.

\section{Model}\label{sec:model}
 We consider an antiferromagnetic Heisenberg model with single-ion anisotropy in the presence of a magnetic field, often used to model the quantum dynamics of magnetic clusters and molecular nanomagnets (MNMs). The nearest-neighbor Hamiltonian we consider is given by
\begin{align}
	\label{eq:Hamiltonian}\hspace{-0.5em}H\hspace{-0.2em}=\hspace{-0.2em}\sum_{\langle i,j\rangle}J_{ij}\boldsymbol{S}_{i}\cdot \boldsymbol{S}_{j}\hspace{-0.2em}+\hspace{-0.2em}\sum_{i=0}^{L-1} \boldsymbol{S}_i\cdot D_i\cdot \boldsymbol{S}_i\hspace{-0.2em}+\hspace{-0.2em}\mu_B  \sum _{i=0}^{L-1}\boldsymbol{B}\cdot g_i\cdot\boldsymbol{S}_i,
\end{align}
where $L$ is the number of spins, $\langle i,j\rangle$ denotes nearest-neighbor pairs $i$ and $j$ and $\boldsymbol{S}_i\equiv(S^x_i,S^y_i,S^z_i)$ is the spin-$S$ angular momentum operator at site $i$ (located in real space at $\boldsymbol{r}_i$) with eigenstates $|S,M_s\rangle$. The first term in the Hamiltonian represents the Heisenberg antiferromagnetic interactions between spins at $i$ and $j$ with exchange constant $J_{ij}$. The second term is the single-ion anisotropy, where $D_i$ is the $3\times 3$ matrix of anisotropy coefficients at site $i$. The third term represents Zeeman interactions, where $g_i$ is a $3\times 3$ matrix representing the anisotropic g-tensor, $\mu_B$ is the Bohr magneton and $\boldsymbol{B}\equiv(B_x,B_y,B_z)$ is the externally applied magnetic field. While the quantum algorithms in this paper can be used to simulate the most general case, we consider the following simplifying assumptions:  $J_{ij}= J$ when $i$ and $j$ are nearest neighbors, $g_i^{\alpha\beta}=g\delta_{\alpha,\beta}$ where g is the Land\'{e} g-factor, and the single ion anisotropy is site-independent and non-zero only along the $z$ axis, $D_i^{zz}=D$, and $D_i^{\alpha\beta}=0$ otherwise.

In the following, we review the basic relationship between the scattering cross-section measured in neutron scattering experiments and the magnetic response function~\cite{Boothroyd2020}. Consider a quantum system being probed with neutrons at zero temperature, i.e., in its ground state, $\ket{\psi_0}$, with energy $E_0$ . The partial differential cross-section (into a solid angle $d\Omega$ and energy difference $dE_f$) measured  in an inelastic neutron scattering experiment is given by~\cite{Boothroyd2020},
\begin{align}
\frac{d^2 \sigma(\boldsymbol{Q},\omega)}{d\Omega dE_f}=\frac{k_{\rm f}}{k_{\rm i}}\left(\gamma r_0\right)^2 \mathcal{I}(\boldsymbol{Q},\omega)\, ,
\end{align}
where $\boldsymbol{k}_{\rm i},\boldsymbol{k}_{\rm f}$ are the momentum wave-vectors for the incident and scattered neutrons where $\boldsymbol{Q}=\boldsymbol{k}_{\rm i}-\boldsymbol{k}_{\rm f}$, $\gamma=-1.913$, $r_0=\frac{1}{4\pi\epsilon_0}\frac{e^2}{m_ec^2}=2.818$ fm is the classical electron radius and $\hbar\omega$ is the energy transferred from the neutron to the target. For unpolarized neutrons, the magnetic scattering intensity is given by,
\begin{align}
	\label{eq:SQdef}	\mathcal{I}(\boldsymbol{Q},\omega)&=\left|\frac{g}{2}F(Q)\right|^2\sum_{\alpha,\beta}(\delta_{\alpha,\beta}-Q_\alpha Q_\beta) \mathcal{S}_{\alpha\beta}(\boldsymbol{Q},\omega)\, ,\\
	\hspace{-0.1in}\mathcal{S}_{\alpha\beta}(\boldsymbol{Q},\omega)&= \sum_p \sum_{\substack{i,j=0\\i\geq j}}^{L-1}\cos(\boldsymbol{Q}\cdot\boldsymbol{r}_{ij})\times\nonumber \\ 
	&\times \bra{\psi_0}\hat{S}_i^\alpha \ket{\psi_p}\bra{\psi_p}\hat{S}_j^\beta \ket{\psi_0}\delta(\Delta E_p-\hbar \omega). \label{eq:Sab}
\end{align}
Here, $\alpha,\beta \in\{x,y,z\}$, $F(Q)$ is the form factor for the ion assuming identical ions, $\boldsymbol{r}_{ij}=\boldsymbol{r}_i-\boldsymbol{r}_j$ is the separation between the ions, $\ket{\psi_p}$ represents the excited states with excitation energy $\Delta E_p=E_p-E_0$, and $p=1,2,\cdots$. Furthermore, in the absence of magnetic field, it can be shown that only non-zero contributions in the sum expression for the scattering intensity, $\mathcal{I}(\boldsymbol{Q},\omega)$  come from the $\mathcal{S}_{\alpha\alpha}(\boldsymbol{Q},\omega) $ terms, $\alpha=x,y,z$.

In the following sections, we discuss classical and quantum TN algorithms to prepare 
(See \cref{fig:schematic}) the ground state $\ket{\psi_0}$ and $P$ low-lying excited states $\ket{\psi_p}$, where $p=1\cdots P-1$. Furthermore, using these states, we design quantum algorithms to compute static correlation functions defined as,
\begin{align}
	M^{\alpha}_{i}(\ket{\psi})&=\bra{\psi}\hat{S}_i^\alpha\ket{\psi},\label{eq:M-def}\\
	C^{\alpha\beta}_{ij}(\ket{\psi})&=\bra{\psi}\hat{S}_i^\alpha\hat{S}_j^\beta\ket{\psi},\label{eq:C-def}
\end{align}
where $\ket{\psi} \in \{\ket{\psi_p}\}$, as well as magnetic dipole transition
matrix elements of the form,
\begin{align}
\label{eq:Ofullgen}O_{ij;p}^{\alpha\beta}=  \langle \psi_0|\hat{S}^{\alpha}_i|\psi_p\rangle \langle \psi_p|\hat{S}^{\beta}_j|\psi_0\rangle\, .
\end{align}
 The overlaps are then used to compute the magnetic response function as defined in \cref{eq:Sab}.

\section{Classical Tensor Network Methods}\label{sec:CTNmethods}
In this section, we review the basic formalism of MPSs in the context of simulating the eigenstates of the Heisenberg model described in \cref{eq:Hamiltonian}.
CTN methods decompose a high-rank tensor in terms of a network of lower-rank tensors with contracted indices. MPSs are CTN ansatzes with a one-dimensional network geometry and are one of the most prevalent tensor network structures for simulating many-body quantum systems. 
Consider a rank-$L$ tensor with $d^L$ parameters represented by $c_{i_0\dots i_{L-1}}$, where $i_j=0,1\cdots d-1$ and $j=0,\cdots L-1$.
The MPS ansatz decomposes this tensor into $L$ rank-3 tensors as,
\begin{align}
c_{i_0\dots i_{L-1}}&=\mathrm{Tr}\left(\mathbb{A}^{[0]i_0}\dots \mathbb{A}^{[L-1]i_{L-1}}\right)\, .
\end{align}
Each individual rank-3 tensor is represented as a matrix, $\mathbb{A}^{[j]i_j}\equiv A^{[j]i_j}_{\alpha \beta}$, where $\alpha,\beta\leq\chi$ are known as \emph{bond indices} with the maximal dimension of any bond index denoted as the \emph{bond dimension} $\chi$, $j=0\cdots L-1$ denotes the ordering of tensor contraction (or matrix multiplication viewing $\mathbb{A}^{[j]i_j}$ as a matrix), $\mathrm{Tr}\left(\bullet\right)$ is the matrix trace summing over the first and last bond index, and $i_j=1\cdots d$ corresponds to the physical indices of the original tensor. Clearly, the MPS description of the tensor contains $Ld\chi^2$ parameters in comparison to the $d^L$ parameters necessary for the full tensor. While any arbitrary tensor lends itself to an exact MPS with a bond dimension scaling exponentially in the tensor rank as $\chi\sim d^{L/2}$, it has been observed empirically that many physically relevant tensors can be described with good fidelity by tensor networks with a bond dimension that is independent or only a polynomial function of $L$. Thus, the MPS ansatz provides a systematic framework for compressing the information present in large datasets through a careful choice of the bond dimension.

In the context of quantum systems, an MPS can be used to describe the coefficient tensor of a quantum wavefunction. For example, the quantum states of $L$ spin-$S$ particles, indexed by their magnetic quantum number $i_\mu$ ($d=2S+1$) can be rewritten as an MPS,
\begin{align}
\label{eq:psiMPS}\hspace{-0.6 em}|\psi\left[A\right]\rangle&=\hspace{-0.8 em}\sum_{i_0\dots i_{L-1}} \hspace{-0.8 em}\mathrm{Tr}\left(\mathbb{A}^{[0]i_0}\dots \mathbb{A}^{[L-1]i_{L-1}}\right)|i_0\dots i_{L-1}\rangle\, ,\hspace{-0.6 em}
\end{align}
where, $i_j=-S,\cdots, S$.
MPSs underlie the celebrated density matrix renormalization group (DMRG) algorithm for one-dimensional quantum systems~\cite{white1992density,Schollwock2011}, and the success of MPSs and their generalizations in describing some strongly correlated quantum systems has connections with quantum information and entanglement theory~\cite{PhysRevLett.100.070502,eisert2013entanglement}.  In the context of DMRG, it is convenient for the purposes of numerical stability to fix the first index of $\mathbb{A}^{[0]}$ and the last index of $\mathbb{A}^{[L-1]}$ to be one-dimensional, which also makes the trace in Eq.~\eqref{eq:psiMPS} unnecessary.  While this is commonly referred to as open boundary conditions in the DMRG literature, we stress that an MPS of this form can also describe a state with periodic boundary conditions, although usually at the expense of increased bond dimension~\cite{verstraete2004density}.

The matrix product structure in Eq.~\eqref{eq:psiMPS}  has a gauge freedom in the definition of the individual matrices, because any full rank $\chi\times \chi$ matrix and its inverse 
can be placed between any two neighboring tensors without changing the physical values.  Using this gauge freedom the MPS representation can be transformed into one of a number of canonical forms~\cite{perez2006matrix,Schollwock2011} which can be numerically obtained straightforwardly using orthogonal decompositions such as the QR decomposition or singular value decomposition (SVD).  A key role will be played in our exposition by MPSs in left-canonical form represented by $\mathbb{L}^{[j]i}\equiv L_{\alpha \beta}^{[j]i}$ satisfying $\sum_{\alpha i }L_{\alpha \beta}^{[j]i\star} L_{\alpha \beta'}^{[j]i}=\delta_{\beta\beta'}$ $\forall j$,  and analogously in the right-canonical form represented by $\mathbb{R}^{[j]i}\equiv R_{\alpha \beta}^{[j]i}$ satisfying $\sum_{\beta i}R_{\alpha\beta}^{[j] i}R_{\alpha' \beta}^{[j] i\star}=\delta_{\alpha\alpha'}$.  Examining the orthonormality relation for the left canonical $[$right canonical$]$ representation, it is clear that reshaping each tensor into a $(\chi d) \times \chi$ $[\chi \times (\chi d)]$ matrix defines an isometric matrix with orthonormal columns (orthonormal rows).  Hence, each of the left or right canonical tensors can be embedded into a $(\chi d) \times (\chi d)$ unitary matrix as,
\begin{align}
\label{eq:Lcembed} L_{\alpha\beta}^{[j] i}&=\langle i \alpha|\hat{U}^{[j]}_{\rm L}|0\beta\rangle\, ,\\
\label{eq:Rcembed} R_{\alpha\beta}^{[j] i}&=\langle i \alpha|\hat{U}^{[j]\dagger}_{\rm R}|0\beta\rangle\, ,
\end{align}
respectively. Thus, this unitary embedding of the MPS tensors may be viewed as a unitary operation acting on a $\chi$-dimensional ancilla state and a $d$-dimensional ``physical" state.  We will return to this point in the \cref{sec:QTN} when we discuss quantum tensor networks, and show how MPS states can be implemented on quantum computers.  
Mixed-canonical form places all tensors to the left of index $j$ in left-canonical form and all tensors to the right of index $j$ in right-canonical form, which makes the tensor at index $j$, which is now called the orthogonality center, the expansion of the wavefunction in an orthonormal basis. We utilize the mixed canonical form to classically optimize the MPS to obtain the ground and excited states.

Analogous to the MPS representation of a quantum state in \cref{eq:psiMPS}, one can define a matrix product operator (MPO) representation of an arbitrary quantum many-body operator, $\hat O$ acting on the same Hilbert space, 
\begin{align}
\hat{O}&=\mathrm{Tr}\left(\hat{\mathbb{W}}^{[0]}\dots \hat{\mathbb{W}}^{[L-1]}\right)\, ,
\end{align}
where, $\hat{\mathbb{W}}^{[j]}\equiv W^{[j]i_j i_j'}_{\alpha \beta}$ on the right hand side can be viewed as bond-indexed ($\alpha,\beta$) matrices whose elements are local operators acting on a single $d$-level system at site $j$, with composition of these local operators understood to be by the tensor product.  As an example, the MPO representation of the operator, $\hat{O}=\sum_i h_i \hat{S}^z_i$ is,
\begin{align}
\hat{\mathbb{W}}^{[0]}&=\left(\begin{array}{cc} h_0 \hat{S}^z_0 &\hat{\mathbb{I}}\end{array}\right)\, ,\\
\hat{\mathbb{W}}^{[0<j<L-1]}&=\left(\begin{array}{cc} \hat{\mathbb{I}}&\hat{{0}} \\ h_j \hat{S}^z_j &\hat{\mathbb{I}}\end{array}\right)\, , \\ 
\hat{\mathbb{W}}^{[L-1]}&=\left(\begin{array}{c} \hat{\mathbb{I}}\\ h_{L-1} \hat{S}^z_{L-1} \end{array}\right)\, ,
\end{align}
in which $\hat{\mathbb{I}}$ and $\hat{0}$ are the $(2S+1)\times (2S+1)$ identity and zero operators, respectively.  Additional details on the representation of common many-body operators as MPOs and MPO arithmetic can be found in Refs.~\onlinecite{mcculloch2007density,wall2012out}.

Next we discuss the classical optimization of the MPSs to obtain the lowest $N$ eigenstates ($\{\ket{\psi_p[A_p]}\}$, $p=0\cdots N-1$) of the Hamiltonian in \cref{eq:Hamiltonian}. The optimization was done using standard DMRG-type variational optimization where the cost function minimizes the expected value of the Hamiltonian $\hat H$ while enforcing orthonormality constraints. The cost functional for the $p^{\rm th}$ eigenstate is given by
\begin{align}
 \langle \psi\left[A\right]|\hat{H}|\psi\left[A\right]\rangle-E\langle \psi\left[A\right]|\psi\left[A\right]\rangle -\sum_{p=0}^{N-2}\lambda_{p} \langle \psi\left[A\right]|\psi_{p}\left[A_{p}\right]\rangle \, , \nonumber
\end{align}
where $\hat{H}$ is an MPO representation of the Hamiltonian, $|\psi\left[A\right]\rangle$ is a variational MPS representation, $E$ is a Lagrange multiplier enforcing state normalization, and $\lambda_{p}$ is a Lagrange multiplier enforcing orthogonalization against the previously found $N-1$ eigenstates.  The functional is minimized over pairs of neighboring tensors containing the orthogonality center, sweeping the pair of sites being optimized across all tensors. Convergence is monitored by the energy variance $\langle \left(\hat{H}-\langle \hat{H}\rangle\right)^2\rangle<\varepsilon_V$.  Following convergence, we further reduce the bond dimension by applying variational compression algorithms~\cite{Schollwock2011}.

\section{Quantum tensor network methods}\label{sec:QTN}
In this section, we investigate quantum tensor network methods for evaluation of the magnetic response function. In \cref{subsec:statepre}, we detail how quantum states represented as MPSs can be mapped to equivalent quantum circuits, and a variational compilation approach for evaluation on near-term quantum computers. Next, in \cref{subsec:meas} we discuss quantum algorithms for measurement of various static and dynamic correlation functions.

\subsection{ MPS state preparation algorithm} \label{subsec:statepre}

\begin{figure}
	\begin{center}
		\includegraphics[width=\columnwidth]{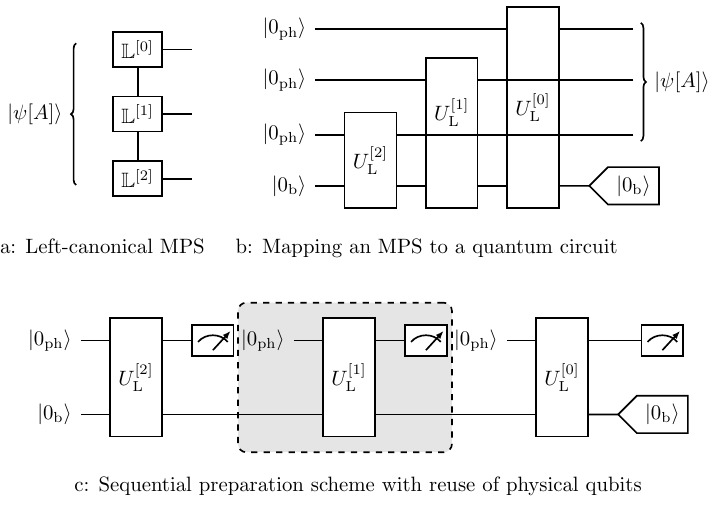}
		\caption{\label{fig:MPSprep} Quantum circuit representation of a MPS. (a) Three-site MPS represented in the left canonical form. (b) Quantum circuit for drawing a sample of an $L=3$ site MPS in the $z$ basis.  The physical qubits are denoted as $\ket{0_{\rm ph}}$ and the bond dimension is spanned by the bond qubits $\ket{0_b}$. (c) Sequential preparation of the circuit presented in (b) using a single physical qudit and MCMR.  The ``unit cell" of the MPS construction and sampling process has been outlined in the dashed box.  The tag on the bond qubit at the far right of the diagram indicates a post-selection of the measurements conditioned on the bond qubits being in the zero state.}
	\end{center}
\end{figure}

The implementation of MPSs on quantum computers rests on the fact that a necessary and sufficient condition for a state to be an MPS on $L$ sites is that it can be sequentially prepared encoding the bond index as a $\chi$-level ancilla, which can be represented using $N_\chi=\log_2(\chi)$  bond qubits, $\{q_b\}$, and the physical index corresponding to  $L$ $d$-level systems (or `qudits"), encoded in $N_d=\log_2(d)$ physical qubits $\{q_{\rm ph}\}$~\cite{schon2007sequential}. More concretely, using the unitary embedding of a left-canonical MPS in Eq.~\eqref{eq:Lcembed}, we can prepare an MPS as
\begin{align}
	\nonumber &\left(\hat{U}^{[0]}_{\rm L}\ket{0^{[0]}_{\rm ph}}\dots \left(\hat{U}^{[L-2]}_{\rm L}\ket{0^{[L-2]}_{\rm ph}}\left(\hat{U}^{[L-1]}_{\rm L} \ket{0^{[L-1]}_{\rm ph}0^{\phantom{[]}}_{\rm b}}\right)\right) \dots\right)\\
	&=\sum_{i_0 \dots i_{L-1}}\mathbb{L}^{[0]i_0}\dots \mathbb{L}^{[L-1]i_{L-1}}|i_0\dots i_{L-1}\rangle\ket{0_b} \nonumber\\
	& = |\psi\left[A\right]\rangle|0_b\rangle \, \label{eq:LCprep}
\end{align}
where, $\ket{0_b}$ is the $\chi$-level bond qudit, $\ket{0^{[\mu]}_{\rm ph}}$ is the $d$-level physical qudit at site $j$ and the unitary operators $\hat{U}^{[j]}_{\rm L}$ act on the Hilbert space spanned by physical qudit at site $j$ and the bond qudit.  The ancilla bond qubits begin in the vacuum and, importantly, decouple from the physical qubits back to the vacuum state at the end of the preparation sequence -- this corresponds to the one-dimensional bond dimensions of the boundary of the first and last tensor ($j=0$ or $L-1$) in the MPS definitions from Sec.~\ref{sec:CTNmethods}. 
In \cref{fig:MPSprep}~(a), we represent an example $L=3$ site MPS state and the corresponding quantum circuit in \cref{fig:MPSprep}~(b) that utilizes $L$ physical qudits and $1$ bond qudit. Thus, a direct mapping of a classical MPS state into a quantum circuit for evaluation on a quantum computer would require $LN_d+N_\chi$ qubits.

Static properties of the quantum state maybe extracted from measurements performed on the physical qudits. The structure of the unitary circuit representation of MPS in \cref{fig:MPSprep}~(b) shows that the physical qubit at site $j$ remains idle following the application of the unitary, $\hat{U}_{\rm L}^{[j]}$. Thus, instead of measuring the qudits at the end of the circuit, static observables may be extracted from measurements on the $j^{\rm th}$ qudit that immediately follow the application of $\hat{U}_{\rm L}^{[j]}$. Going a step further, we can reduce the qudit requirements further  by resetting the measured physical qudit ($j^{\rm th}$ qudit) and reusing it to represent as the $(j-1)^{\rm th}$ physical qudit.
Thus, through mid-circuit measurement and reuse of the physical qubits the qubit resources necessary to implement the state preparation algorithm reduces to $N_\chi+N_d$. We refer to this scheme as a sequential preparation of a quantum tensor network state. In \cref{fig:MPSprep}~(c), we show an example of such a measurement for the 3-site MPS state. This logarithmic dependence on the bond dimension $\chi$ and the independence of the requirement on the length of the physical system $L$ point to a significant resource advantage over the resources required to represent the same state as a classical MPS.  In Fig.~\ref{fig:MPSprep} we identify a ``unit cell" of the sequential preparation using the dashed box -- this unit cell defines an elemental increment of the algorithm across a lattice site. To keep future diagrams uncluttered, we will present only this unit cell with the understanding that it is to be repeatedly applied in a sequential preparation procedure.

While we have presented the MPS preparation scheme starting from the left-canonical form of the MPS, it is straightforward to also prepare the state using the unitary embeddings of the right-canonical tensors defined in Eq.~\eqref{eq:Rcembed}, as
\begin{align}
\nonumber &\left(\hat{U}^{[L-1]\dagger}_{\rm R}\ket{0^{[L-1]}_{\rm ph}}\dots \left(\hat{U}^{[1]\dagger}_{\rm R}\ket{0^{[1]}_{\rm ph}} \left(\hat{U}^{[0]\dagger}_{\rm R} \ket{0^{[0]}_{\rm ph}0_b^{\phantom{[]}}}\right)\right)\dots\right) \\
&=\sum_{i_0 \dots i_{L-1}}\mathbb{R}^{[0]i_0}\dots \mathbb{R}^{[L-1]i_{L-1}}|i_0\dots i_{L-1}\rangle |0_a\rangle \nonumber \\
&= |\psi\left[A\right]\rangle |0_a\rangle \, \label{eq:RCprep}.
\end{align}
From the above expression for preparation of the MPS state, it is clear that the unitary-embedded right-canonical tensors can also be employed as an {\it un-preparation} sequence,
\begin{align}
\label{eq:URundo}\hat{U}^{[0]}_{\rm R} \hat{U}^{[1]}_{\rm R}\dots \hat{U}^{[L-1]}_{\rm R}|\psi\left[A\right]\rangle |0_a\rangle=|0_0\dots 0_{L-1}\rangle|0_a\rangle \, .
\end{align}
We will return to the un-preparation sequence in \cref{subsec:meas}, when we use it to compute overlaps between MPSs in the sequential preparation scheme.

\subsubsection*{Compilation of MPS operations to quantum hardware}\label{subsubsec:compilation}

Having established the state-preparation algorithm for an arbitrary classical MPS, its implementation  on quantum hardware rests on compilation of the unitary embeddings to a native gate-set. We adapt 
 a greedy compilation procedure previously demonstrated in Refs.~\onlinecite{wall2020Generative,wall2021tree,WallPRA2022} to find variationally the unitary embeddings (in a given gate-set) of isometric MPS tensors provided in the left or right canonical form. The parameters of the variational circuit are optimized with respect to the $L_2$ distance between the unitary operator and the isometric matrix  in the region where the latter has support with a cost function given by
\begin{align}
\label{eq:varcostfunc}\mathcal{C}\left(\hat{U},\mathbb{L}^{[i]}\right)&=\sum_{(a,b)\in \mathcal{S}}\left|\hat{U}_{ab}-L^{[i]}_{ab}\right|^2\, .
\end{align}
Here, $L^{[i]}$ denotes the matrix representation of the reshaped rank-3 tensors $\mathbb{L}^{[i]}$ [See discussion above \cref{eq:Lcembed,eq:Rcembed}] and $\mathcal{S}$ denotes the set of indices in the matrix such that the isometry elements are nonzero to within some tolerance $\delta$, i.e. $|L^{[i]}_{ab}|>\delta$.  If this cost function is less than a user-defined tolerance $\varepsilon_C$, we accept the unitary embedding and set it as $\hat{U}^{[i]}$. The choice of the tolerances $\delta$ and  $\varepsilon_C$ directly determine the total gate count of the compilation.

\begin{figure}[t]
	\begin{center}
  \includegraphics[width=\columnwidth]{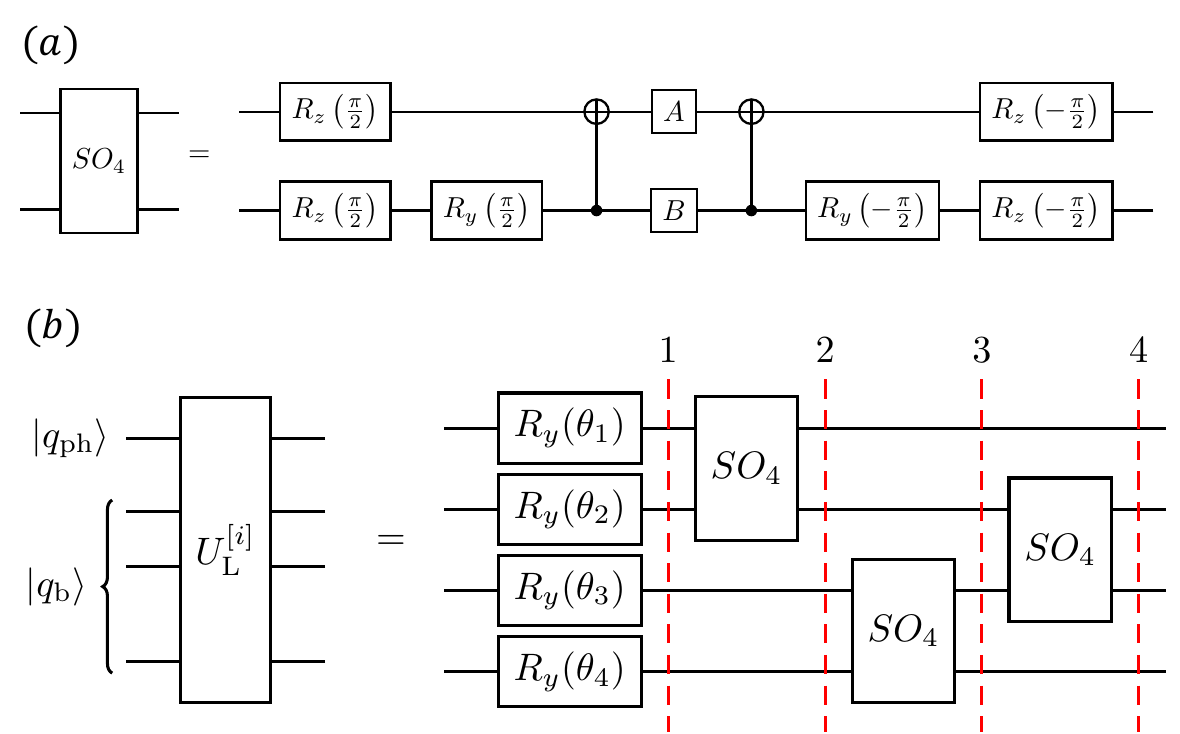}  
  \caption{\label{fig:SO4} (a) Circuit for implementing a general element of SO(4).  Here, $A$ and $B$ are parametrized by SU(2) unitary operators that are variationally optimized. (b) An example variational circuit for a MPS state preparation unitary after 4 steps of the optimization of the cost function in \cref{eq:varcostfunc}. 
  }
  \end{center}
  \end{figure}

The variational circuit is built iteratively increasing in circuit depth, with construction of the circuit halting when the $\mathcal{C}<\varepsilon_C$. We begin with an ansatz consisting of a single-qubit rotation on each qubit and optimize the parameters for the cost function. If the tolerance isn't achieved, we construct a new set of candidate circuits from the previous one by appending a single entangling gate, with the candidates distinguished by the particular entangling gate and the pair of qubits it connects. We parametrize the entangling gate as the general SO(4) gate whose circuit ansatz is shown in Fig.~\ref{fig:SO4}~(a)~\cite{vatan2004optimal}. 
Each of the candidate circuits is optimized with respect to the cost function and their final cost functions recorded.  If any gate meets the cost function tolerance that circuit is selected as the unitary embedding, otherwise the $h$ lowest cost function gates form the starting point for a new set of candidate gates, where $h$ is a user-defined parameter.  As before, the candidate gates are constructed by adding a single entangling gate, and the optimization procedure is repeated until convergence. An example of this step by step construction is provided in \cref{fig:SO4} (b). This process identifies short-length unitary embeddings of MPS tensors without being overly sensitive to noise resulting from DMRG optimization, and can be readily adapted to include restrictions on gate sets, hardware topology, noise, or other NISQ-relevant device characteristics. We note here that there is potentially a wide variety of other approaches that could be applicable to this variational compilation problem. Nonetheless, in practice, we find that our approach towards the variational compilation of the tensors does provide constant factor improvements over exact compilation methods, as is discussed further in \cref{sec:results}. For the purposes of this comparison, we utilize the native compilation tool available through the Qiskit library to obtain the exact compilation of the isometric matrices~\cite{Qiskit}. Note that the cost to exactly compile the isometry increases exponentially with the bond-dimension making it impractical to implement for large bond-dimension MPS.
  
The variational approach just described requires a classical MPS representation to generate an equivalent QTN representation. Evaluation of the cost function scales exponentially with the number of bond qubits and therefore, is intractable for systems that are not classically simulable. In this work, our focus is on developing efficient algorithms to extract static and dynamic correlation functions and therefore we only consider systems where a classical representation of the MPS has already been evaluated. In practice, we anticipate that the classical MPS (compiled to a QTN) with a small bond dimension will serve as a starting point for a variational search of quantum MPS states by including additional bond qubits and parameterized gate sequences, potentially pushing the QTN to towards a classically intractable regime.  Thus, the process outlined here may be viewed as a way to find an efficient classical pre-conditioner for identifying well-performing unitary embeddings of MPSs of modest bond dimension that can then be used as variational circuit ans\"{a}tze in their own right. Furthermore, this variational approach to compilation may also provide a certain degree of robustness to gate errors, especially coherent errors, as is typically observed in variational quantum algorithms~\cite{Cerezo_2021,quiroz2021,Berberich2024,Biamonte2024}.

\subsection{Measurement of observables}\label{subsec:meas}

Let us now discuss algorithms for extracting relevant physical quantities from the MPSs prepared on a quantum computer.  We discuss the procedure to use the state-preparation circuit for extracting (in expectation) the following quantities from the encoded MPS,
\begin{enumerate}
	\item {\it Static observables}-- We calculate single and two-point functions such as magnetization, $M^\alpha_i$ and spin-spin correlations, $C^{\alpha\beta}_{ij}$ [defined in \cref{eq:C-def,eq:M-def}], which can then be used to evaluate other observables of interest such as excitation energies.
	\item {\it Wavefunction Overlap}-- We discuss a sequential algorithm to compute the overlap between two MPS states, $\ket{\psi[A_0]}$ and $\ket{\psi[A_p]}$ given by $\left|\braket{\psi[A_p]|\psi[A_0]}\right|^2$.
	\item {\it Magnetic dipole transition matrix elements}-- We discuss two approaches to extract the magnetic dipole transition matrix elements, $O_{ij;p}^{\alpha\beta}$ defined in \cref{eq:Ofullgen}. 
\end{enumerate}

\subsubsection{Static Observables}\label{subsubsec:static-obs}
\begin{figure}
	\begin{center}
		\includegraphics[width=\columnwidth]{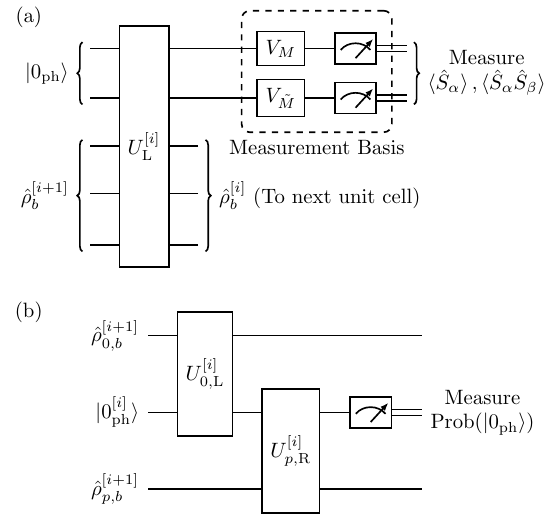}  
		\caption{\label{fig:OverlapSequentialUC} Unit cell for estimation of static observables in a given MPS and wavefunction overlap between two MPSs. (a) The MPS $|\psi\left[A\right]\rangle$ is prepared in the left canonical form, and the measurement basis for physical qubits on each site (denoted by $V_M$ and $V_{\tilde{M}}$) are chosen according to the observable of interest. The measurement statistics are then correlated to extract the expectation value of the observable. (b) Two MPSs, $|\psi\left[A_0\right]\rangle$ with left-canonical unitary completions $\{\hat{U}^{[i]}_{0,L}\}$ and $|\psi\left[A_p\right]\rangle$ with right canonical unitary completions $\{\hat{U}^{[i]}_{p,R}\}$.  State preparation of $|\psi\left[A_0\right]\rangle$ across a single site is applied to physical qudit $|0_j\rangle$ using the ancilla $|\psi_{a,0}\rangle$ by $\hat{U}_{\rm L}$, state ``un-preparation" of this site according to $|\psi\left[A_p\right]\rangle$ is applied to physical qudit $|i_j\rangle$ using the ancilla $|\psi_{a,p}\rangle$ by $\hat{U}_{\rm R}$, and the physical qudit is measured and reset.  The fraction of shots where all physical measurements and measurements of bond qubits at the end of the sequence yield the zero state is the squared overlap $\left|\langle \psi\left[A_p\right]|\psi\left[A_0\right]\rangle\right|^2$.}
	\end{center}
\end{figure}

Static or equal-time correlation functions are obtained from the circuit by combining measurement outcomes in different basis states of the physical qubits in the circuit. Measurement of one-and two point correlation functions can be simply measured by rotating the basis of the physical qudits before measurement and post-processing samples. This procedure is schematically shown in Fig.~\ref{fig:OverlapSequentialUC}~(a).

The simplest example, is that of spin-1/2 models, the physical index is encoded in a single qubit, and we identify the spin operators with the Pauli matrices, 
\begin{align}
	\hat{S}^\alpha_j\rightarrow \frac{1}{2}\sigma^\alpha_j
\end{align}
where, $\alpha=x,y,z$ and $\hat{\sigma}^\alpha_j$ represents the Pauli matrix. In this case, it is straightforward to extract the magnetization and correlations 
$M^{\alpha}_{i}=\frac{1}{2}\langle\hat{\sigma^\alpha_i}\rangle$ and $C^{\alpha\beta}_{ij}=\frac{1}{4}\langle \hat{\sigma}_i^\alpha\hat{\sigma}_j^\beta\rangle$ from the measurement statistics in the appropriate basis [set by the operation $V_M$ in \cref{fig:OverlapSequentialUC} (a)]. For example, consider the case when $\alpha, \beta=x$. In this case the measurement basis is in the $x$ direction, and thus in the sequential scheme Eq.~\eqref{eq:LCprep}, we must apply a Hadamard gate on the physical qubit after preparation and before measurement.  Each measurement bitstring constitutes a sample in the $x$ basis $\left(x_0,\dots,x_L\right)_m$ where $m$ is a shot index, and so we can estimate $\langle \hat{\sigma}^x_i\rangle =\sum_{m} \left(1-2\delta_{x_{i,m},1}\right)/M$ and $\langle \hat{\sigma}^x_i \hat{\sigma}^x_j\rangle  = \sum_{m} \left(1-2\delta_{x_{i,m},1}\right)\left(1-2\delta_{x_{j,m},1}\right)/M$, where $M$ is the total number of shots. Utilizing the correlation functions along $x$, $y$, $z$ basis states, it is possible to estimate the expectation value of the Hamiltonian that estimates the energy of the state.

More general spin models are encoded using multiple physical qubits. For example, a single spin-3/2 has a four dimensional Hilbert space and thus needs two physical qubits to encode. In this case, we can map the spin operators to sums of tensor products of two Pauli operators,
\begin{align}
	\hat{S}^x &= \frac{\sqrt{3}}{2}\hat{I}\hat{\sigma}^x + \frac{1}{2}\hat{\sigma}^x\hat{\sigma}^x + \frac{1}{2}\hat{\sigma}^y\hat{\sigma}^y\label{eq:sx-3/2}\\
	\hat{S}^y &= \frac{\sqrt{3}}{2}\hat{I}\hat{\sigma}^y + \frac{1}{2}\hat{\sigma}^y\hat{\sigma}^x - \frac{1}{2}\hat{\sigma}^x\hat{\sigma}^y\label{eq:sy-3/2}\\
	\hat{S}^z &= \frac{1}{2}\hat{I}\hat{\sigma}^z + \hat{\sigma}^z\hat{I}\label{eq:sz-3/2}
\end{align}
In \cref{subsec:extensions}, we use the above expressions, to determine the basis for measurements necessary for extracting measurement statistics to compute observables of spin-3/2 operators such as the magnetization and correlation functions. It is clear that in this case one needs to average measurement results from multiple different measurements to extract observables.

\subsubsection{Wavefunction Overlap}\label{subsubsec:overlap}

 In this section we discuss an algorithm for estimating the overlap $\left|\langle \psi\left[A_p\right]|\psi\left[A_0\right]\rangle\right|^2$ between two states represented as MPSs.  In this sequential preparation scheme, the overlap can be computed by first preparing $\ket{\psi[A_0]}$ [See \cref{eq:LCprep}] followed by an un-preparation of $\ket{\psi[A_p]}$ [See \cref{eq:URundo}], and calculating the probability of having all the physical and bond qubits return to the `$0$' state.  A diagram of the sequential preparation unit cell for evaluating the overlap is given in Fig.~\ref{fig:OverlapSequentialUC}~(b). More specifically, let us denote $\hat{\mathcal{L}}_0\equiv \hat{U}^{[0]}_{0,L}\dots \hat{U}^{[L-2]}_{0,L} \hat{U}^{[L-1]}_{0,L}$ to be the state preparation sequence for $|\psi\left[A_0\right]\rangle$ from Eq.~\eqref{eq:LCprep} and $\hat{\mathcal{R}}_p$ to be the sequence $\hat{U}^{[0]}_{p,R} \hat{U}^{[1]}_{p,R}\dots \hat{U}^{[L-1]}_{p,R}$ from Eq.~\eqref{eq:URundo} which takes $|\psi\left[A_p\right]\rangle$ into $|0_0\dots 0_{L-1}\rangle|0_a\rangle$.  Clearly, we have the following identity,
\begin{align}
	&\left|\langle \psi\left[A_p\right]|\psi\left[A_0\right]\rangle\right|^2=\nonumber\\
	&\quad\quad\quad\left|\bra{0_a}\bra{0_0\cdots0_{L-1}}\hat{\mathcal{R}}_p\hat{\mathcal{L}}_0\ket{0_0\dots 0_{L-1}}\ket{ 0_a}\right|^2.
\end{align}
 Thus, the probability $\left|\langle \psi\left[A_p\right]|\psi\left[A_0\right]\rangle\right|^2$ can be estimated from the fraction of all zero bitstrings obtained as the result of measurement of all physical and bond qubit registers in the computational basis following application of $\hat{\mathcal{R}}_p\hat{\mathcal{L}}_0$.  The bond qubits end up in the $|0\rangle$ state by construction of the sequential preparation process of the MPS, and so this post-selection on the bond qubits will occur with high fidelity given that the unitary embedding of the MPS tensors is accurate. The accuracy of this embedding is controlled in our greedy compilation process through the tolerance $\varepsilon_C$.  Further, it should be noted that this overlap can be estimated using only $\log_2\chi_0+\log_2\chi_p+\log_2 d$ qubits, where $\chi_0$ and $\chi_p$ correspond to the bond dimension necessary for representing $\ket{\psi\left[A_0\right]}$ and $\ket{\psi\left[A_p\right]}$ using sequential applications of the $\hat{U}_{0,L}$ and $\hat{U}_{p,R}$ unitary completions respectively.   We stress that the use of sequential preparation requires that $|\psi_0\rangle$ be compiled in left-canonical form and $|\psi_p\rangle$ be compiled in right-canonical form.  It can be shown that this algorithm for overlap, which we call the adjoint method, has reduced variance for a fixed number of shots compared with a standard SWAP test~\cite{EAQC2}.  
 
 We note here that the conditions of left-and right-canonical form for a QTN representation of MPS are accomplished by the order of operations applied to the combined physical and ancilla register, see Eq.~\eqref{eq:LCprep}, as operations on a quantum computer are naturally unitary and thus, the isometric condition is automatically satisfied.  Hence, expansion of a quantum MPS in one of these canonical forms can be achieved by modifying the unitaries while preserving the ordering of operations.  Additional care is needed to ensure the preservation of open boundaries in finite-size simulations.  An additional complication comes when we wish to expand a left-canonical representation and a right-canonical representation of the same state consistently, as the circuit representations of the two have a complex relationship.  We note that the above algorithm provides a means of doing so variationally, where the overlap between the modified circuits can be maximized with respect to their parameters.  A similar construction in a related context for infinite systems was given in Ref.~\cite{PhysRevResearch.5.033187}.

\subsubsection{Magnetic dipole transition matrix elements}\label{subsubsec:diople-elements}

For the purpose of computing spectral functions of the form Eq.~\eqref{eq:SQdef}, we require the ability to compute the expected value of the energy, as well as magnetic dipole transition matrix elements of the form,
\begin{align}
O_{ij;p}^{\alpha\beta}\equiv  \langle \psi_0|\hat{S}^{\alpha}_i|\psi_p\rangle \langle \psi_p|\hat{S}^{\beta}_j|\psi_0\rangle\, ,
\end{align}
where, $\ket{\psi_p}$ denotes the $p^{\rm th }$ excited state. First, we discuss an algorithm to calculate this overlap for the special case when $\alpha=\beta$ for translationally invariant systems. Following that, for the more general case ($\alpha\neq\beta$, arbitrary quantum states), we outline a strategy involving a SWAP test~\cite{Buhrman2001_swaptest}. The advantage of the former method is in the reduced qubit resource requirements as compared to the SWAP test which we detail below.

{\it Translationally Invariant systems}--
  The elements defined in \cref{eq:Ofullgen}  for the case $\alpha=\beta$ and  when the system is translationally invariant, simplify to a function only of the distance between the sites $i$ and $j$, i.e.  $O_{ij;p}^{\alpha\alpha}\sim f\left((j-i)\mathrm{mod}L\right)$.
    As a result, we can obtain these matrix elements by taking advantage of Fourier transforms. Defining the Fourier spin operators as
  \begin{align}
  \label{eq:Sfourier}\hat{\tilde{S}}^{\alpha}_{k}&\equiv \mathcal{F}_k\left[\left\{\hat{S}^{\alpha}_j\right\}\right]=\sum_{j=0}^{L-1}e^{2\pi i j k/L}\hat{S}^{\alpha}_j\, .
  \end{align}
  where $k=0\cdots L-1$, we can transform the expression for $O^{\alpha\alpha}_{i,j;p}$ as
\begin{align}
\label{eq:OFourier}O_{0,0+j;p}^{\alpha\alpha}&=\mathcal{F}^{-1}_j\left[ \left\{\left|\bra{\psi_p}\hat{\tilde{S}}^{\alpha}_{k}\ket{\psi_0}\right|^2\right\}\right]\, ,
\end{align}
in which $\mathcal{F}^{-1}_j\left(\bullet\right)$ denotes the inverse Fourier transform.
Hence, to compute the spectral function in the translationally invariant case we only need to calculate the square modulus matrix elements $\left|\bra{\psi_p}\hat{\tilde{S}}^{\alpha}_{k}\ket{\psi_0}\right|^2$ for $k=0,1,\dots,L-1$. 

\begin{figure*}[t]
	\begin{center}
  \includegraphics[width=\linewidth]{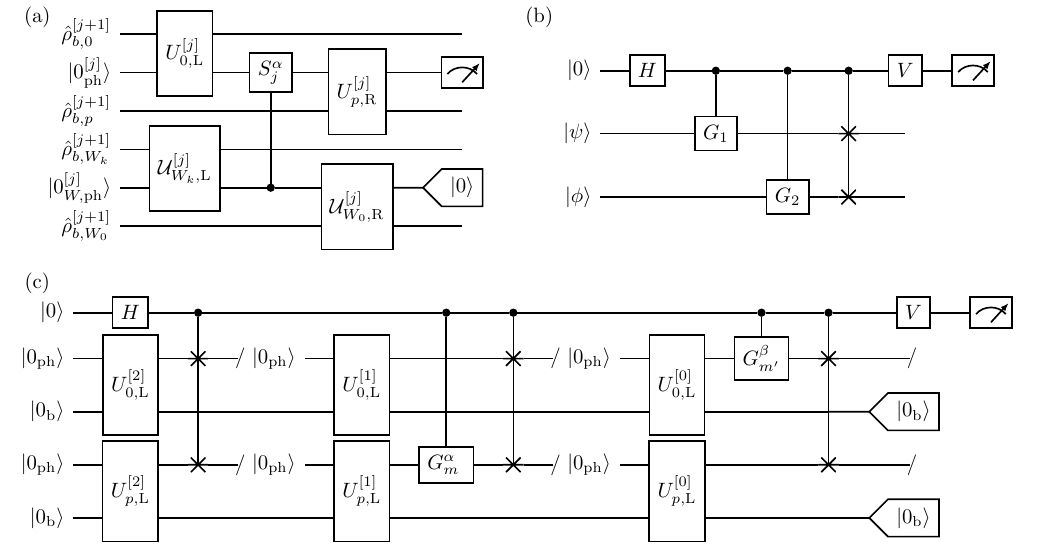}  
  \caption{\label{fig:SpectralSequentialUC} (a) Unit cell for estimation of the magnetic dipole transition matrix elements for the translationally invariant spin-half system. The circuit evaluates the absolute value of the matrix element of the operator $\hat{\tilde{S}}_k^{\alpha}$ between two states $|\psi\left[A_0\right]\rangle$ with left-canonical unitaries $\{\hat{U}^{[j]}_{0,L}\}$ and $|\psi\left[A_p\right]\rangle$ with right canonical unitaries $\{\hat{U}^{[j]}_{p,R}\}$.  The unitaries $\hat{\mathcal{U}}^{[j]}_{W_k,L}$ and $\hat{\mathcal{U}}_{W_0,R}$ encode the MPS representation of the states $|W_k\rangle$ and $|W_0\rangle$ in left- and right-canonical form respectively. (b)  SWAP test strategy for measuring the real or imaginary part of $\langle \psi| \hat{G}_2 |\phi\rangle\langle \phi|\hat{G}_1|\psi\rangle$ given registers prepared with $|\psi\rangle$ and $|\phi\rangle$  and two unitary operations $\hat{G}_1$ and $\hat{G}_2$. The real/imaginary part correspond to the choice of measurement basis for the ancilla: $\hat{V}\equiv$ Hadamard or $\hat{R}_x\left(-\frac{\pi}{2}\right)$ respectively.(c) General SWAP test strategy for measuring the real part of $\langle \psi_0|\hat{G}^{\alpha}_{1,m}|\psi_p\rangle\langle \psi_p|\hat{G}^{\beta}_{0,m^\prime}|\psi_0\rangle$ for $L=3$ site MPSs $|\psi_0\rangle$ and $|\psi_p\rangle$ with unitary embeddings $\{\hat{U}^{[j]}_{0,L}\}$ and $\{\mathcal{U}^{[j]}_{p,L}\}$, respectively. $\hat{V}$ is chosen in the same manner as in (b).
   }
  \end{center}
  \end{figure*}

Now,  we can adapt the sequential algorithm for computing overlaps from \cref{subsubsec:overlap} to evaluate the overlap in \cref{eq:OFourier}. We just need to prepare a state $\ket{\tilde{\psi}_0}\propto\hat{\tilde{S}}_k^{\alpha}|\psi_0\rangle$ before taking the overlap with $|\psi_p\rangle$.  However, as can be seen from \cref{eq:Sfourier}, the operator $\hat{\tilde{S}}_k^{\alpha}$  is not a simple tensor product of local operators, and therefore, its application cannot be achieved simply by acting with local operators on the physical qubit registers before the measurement.  We can circumvent this issue by noting that $\hat{\tilde{S}}_k^{\alpha}$ can be written exactly as an MPO with bond dimension 2 with the MPO matrix representation
\begin{align}
\label{eq:SkMPO}\hat{\mathbb{W}}^{[0<j<L-1]}&=\left(\begin{array}{cc} \hat{\mathbb{I}}&\hat{{0}} \\ e^{2\pi i k j/L} \hat{S}^{\alpha}_j &\hat{\mathbb{I}}\end{array}\right)\, .
\end{align}
For systems where the physical qubits are spin-half systems ($\hat{S}^\alpha_j\propto \sigma^\alpha_j$), we can leverage the above representation to prepare the state $\ket{\tilde{\psi}_0}$ by the following steps.
\begin{enumerate}
	\item Prepare the MPS state $\ket{W_k}$ on some ancillary qubits, defined as
	\begin{align}
		|W_k\rangle&=\frac{1}{\sqrt{L}}\sum_{j=0}^{L-1}e^{2\pi i j k/L} |0\dots 1_j \dots 0\rangle \, ,
	\end{align}
	where the qubits denoted in the sum correspond to the physical qubits necessary to prepare the $\ket{W_k}$ state. Crucially, this state has an analytical representation as an MPS with bond dimension 2 with a known gate decomposition~\cite{wall2020Generative}, and so the state preparation and adjoint operations can be performed sequentially with an additional bond qubit. We represent this state preparation operation by the set of unitary embeddings of the left-canonical form given as $\{\hat{\mathcal{U}}_{W_k,L}^{[j]}\}$
	\item For each unit cell (denoted by physical site $j$), apply a controlled $\hat{S}^{\alpha}$ operation on the pair of physical qubits ($\ket{q}_{\rm ph}$,$\ket{q}_{W,{\rm ph}}$) with the control on the $W$ state physical qubit.  The resulting state is
	\begin{align}
			\frac{1}{\sqrt{L}}\sum_{j=0}^{L-1}e^{2\pi i j k/L} |0\dots 1_j \dots 0\rangle \hat{S}^{\alpha}_j |\psi_0\rangle\, .
	\end{align}
	where the controlled operation can again be applied in a sequential manner.
	\item  Decouple the physical qubits used to prepare the state $|W_k\rangle$ by applying the adjoint of the preparation procedure (or using the un-preparation procedure) with $k=0$ to the register $|i_0'\dots i_{L-1}'\rangle$, which results in the state
	\begin{align}
		 &\ket{0\dots 0}\otimes \left[\frac{1}{L} \sum_{j=0}^{L-1}e^{2\pi i j k/L} \hat{S}^{\alpha}_j \ket{\psi}\right] \nonumber \\
		= &\ket{0\dots 0}\otimes\frac{1}{L} \hat{\tilde{S}}_k^{\alpha} |\psi\rangle\propto\ket{\tilde{\psi}_0}\, ,
	\end{align}
	as was desired. This can be achieved by the right-canonical unitary embedding $\{\hat{\mathcal{U}}_{W_0,R}^{[j]}\}$ for the state $\ket{W_k}$ followed by a measurement of the physical qubit that is post-selected to be in the state $\ket{0}$. 
\end{enumerate}

This method of conditioning operations on an appropriately defined, entangled ancilla register that is then disentangled and decoupled can be thought of as the QTN version of an MPO.  Using the state generated by the steps above, we can now estimate the overlap by un-preparing the state $\ket{\psi_p}$ with its right-canonical embeddings, $\{\hat{U}^{[j]}_{p,R}\}$. Thus, we find the unit cell for the complete estimation of $\left|\bra{\psi_p}\hat{\tilde{S}}^{\alpha}_{k}\ket{\psi_0}\right|^2$ by sequential preparation in Fig.~\ref{fig:SpectralSequentialUC}. As with the state overlap estimation in \cref{subsubsec:overlap}, the squared matrix element is estimated from the fraction of shots in which all physical qubit measurements yield zero conditioned on all bond qubits being in the zero state at the end of the sequence.  Again, the conditioning on the bond qubits yields very little overhead in practice for MPSs which are faithfully compiled into unitaries, and in particular the post-selection on the $|W_k\rangle$ states will occur with high fidelity given the analytic representation of that MPS as a quantum circuit.  The qubit resource requirements for this algorithm are $\sim\log_2\chi_0+\log_2\chi_p+4$ (for spin-1/2 systems), again independent of the number of lattice sites in the system.  Finally, we note that this method of preparing non-local many-body operators using MPS state preparation and un-preparation interspersed with conditional operations can be employed in other settings, such as with other variational quantum eigensolver ans\"{a}tze.  We also expect that this approach can be generalized to other, more complex MPO representations, and so may find greater use within the quantum tensor network community.

{\it Arbitrary Spin Hamiltonians}--Now, let us present an algorithm for the measurement of the overlaps $O_{ij;p}^{\alpha\beta}$ in Eq.~\eqref{eq:Ofullgen} for the most general case with  arbitrary spins and $\alpha\neq\beta$, i.e. does not require translational invariance.  It is based on the canonical SWAP test for measuring overlaps of two unknown quantum states utilizing a Controlled-SWAP (CSWAP) gate and phase kickback to recover the real or imaginary parts of the overlaps~\cite{Buhrman2001_swaptest}.  

 Let us start by reviewing an adaptation of the SWAP test that would allow us to evaluate a matrix element of the form,
 \begin{align}
	\mathcal{M}&=\bra{\psi}\hat{G}_2\ket{\phi}\bra{\phi}\hat{G}_1\ket{\psi}.
 \end{align}
 where $\hat{G}_{1,2}$ are unitary operators that act on the quantum states $\ket{\psi}$ and $\ket{\phi}$.
 In the general case where $|\psi\rangle$ and $|\phi\rangle$ are given in separate registers as input to the algorithm, we consider the circuit shown in Fig.~\ref{fig:SpectralSequentialUC}~(b).  Here, SWAP operations and local operations conditioned on an ancilla qubit prepared in the $\ket{+}$ state are applied. It can be verified by direct computation that measurement of this ancilla qubit in the $\ket{\pm}$ basis (by applying a Hadamard gate before measurement) yields the zero state with probability,
\begin{align}
P_0&=\frac{1}{2}\left[1+\mathrm{Re}\left(\mathcal{M}\right)\right]\, ,
\end{align}
where $\mathrm{Re}\left(\bullet\right)$ denotes the real part.  By modifying the choice of the measurement basis of the ancilla, [changing the Hadamard to a $\hat{R}_x(-\pi/2)$] it can be verified that the output probability of the ancilla qubit is related to the imaginary part of $\mathcal{M}$.

Now, let us discuss the adaptation of the swap-test algorithm to compute the magnetic dipole matrix elements, $O^{\alpha_\beta}_{ij;p}$ given by \cref{eq:Ofullgen} for arbitrary spin systems. We first recognize that the the Hermitian spin operators $\hat{S}^\alpha_j$ can always be written as a linear combination of unitaries,
\begin{align}
	\hat{S}^\alpha_j =\sum_m u_m\hat{G}^{\alpha}_{j,m}\label{eq:S-genexpansion}
\end{align}
where $m$ indexes a complete basis of unitaries (e.g. tensor products of Pauli operators) that spans the spin-qudit Hilbert space; such an expansion for the spin-3/2 operators is provided in \cref{eq:sx-3/2,eq:sy-3/2,eq:sz-3/2}. Now, we can compute the value of $O^{\alpha_\beta}_{ij;p}$ by simply extracting the value of each $\bra{\psi_0}\hat{G}^{\alpha}_{i,m^\prime}\ket{\psi_p}\bra{\psi_p}\hat{G}^\beta_{j,m}\ket{\psi_0}$ and recombining the results from the expansion in \cref{eq:S-genexpansion}. The final step is to adapt the SWAP test for evaluating each individual matrix element to the sequentially prepared MPS case. We note that the SWAPs can be employed on each physical qudit pair in sequence, and so sequential preparation using MCMR can again be utilized.  As an example, in \cref{fig:OverlapSequentialUC} (c) we provide the circuit to extract the real part of $\langle \psi_0|\hat{G}^{\alpha}_{1,m}|\psi_p\rangle\langle \psi_p|\hat{G}^{\beta}_{0,m^\prime}|\psi_0\rangle$ for $L=3$ MPSs, with $|\psi_0\rangle$ and $|\psi_p\rangle$ \emph{both} in left-canonical form, with unitary tensor embeddings $\hat{U}^{[j]}_{0,L}$ and $U^{[j]}_{p,L}$, respectively.  Using sequential preparation and MCMR, the scheme requires $\log_2\chi_0+\log_2\chi_p+2\log_2 d+1$ qubits. In comparison to the translationally invariant circuits, this approach does require the CSWAP gate, which can substantially increase the circuit depth necessary to evaluate the overlaps.  As noted in Sec.~\ref{subsec:statepre}, an MPS representation is a necessary and sufficient condition for a sequential preparation scheme with a $\chi$-level ancilla.  Hence, any variational state methodology that achieves our same qubit resource scaling can be put in the form of a quantum MPS.

\section{Simulation of Magnetic Systems} \label{sec:results}
	In this section, we discuss two example systems to demonstrate the generality of the methods described in previous sections. In the first example, we consider the isotropic, spin-half Heisenberg chain on a ring. Secondly, we consider the spin-3/2 ring, with model parameters that reflect experimental measurements on the Cr$_8$ molecule~\cite{Baker2012}.

	\subsection{Spin-1/2 Heisenberg Chain}\label{subsec:spin 1/2}

	\begin{figure}[t]
		\begin{center}
			\includegraphics[width=\linewidth]{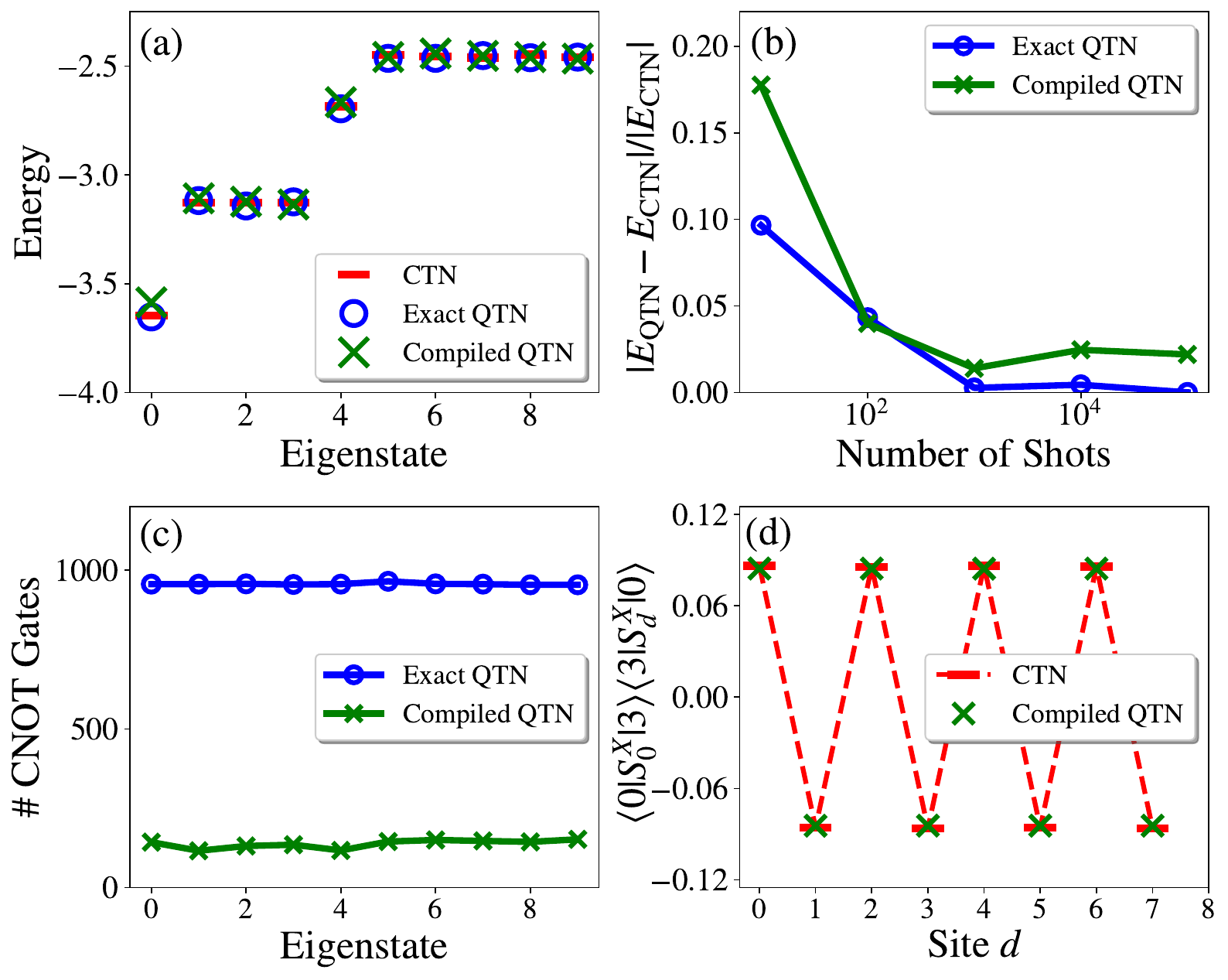}  
			\caption{\label{fig:zeroB-spinhalf} Comparing CTN and QTN methods for evaluating observables in the spin-half Heisenberg chain in zero magnetic field. (a) First 10 Energy levels, $n_{\mathrm{shots}}=10^{4}$.(b) Relative Error between QTN and TN calculations of the ground state energy as a function of $n_{\mathrm{shots}}$. (c) Comparison of circuit depth for compilation of unitary embedding of QTN tensors. An exact compilation (using standard non-variational routine from Qiskit~\cite{Qiskit}) routine is compared with a variational routine discussed in \cref{subsubsec:compilation}. The variational compilation provides a constant factor speed-up in gate counts (characterized here by the number of CNOT gates) when compared to more exact compilation methods. (d) Comparison of Magnetic dipole transition matrix element, $\bra{\psi_0}\hat{S}_0^{x}\ket{\psi_3}\bra{\psi_3}\hat{S}^x_d\ket{\psi_0}$ between QTN and CTN methods. }
		\end{center}
	\end{figure}

	We consider a spin-1/2 Heisenberg chain with periodic boundary conditions and the parameters  $J=1$ and $D = 0$  in the definition of the Hamiltonian provided in \cref{eq:Hamiltonian}. Using a chain of $L=8$ spins, we compute the average energy and dipole transition elements. In order to benchmark the QTN methods, we first extract the $10$ lowest energy eigenstates and eigenvalues for the Hamiltonian using conventional CTN methods as described in \cref{sec:CTNmethods} with the maximum bond dimension set to be $\chi=8$. Comparing with exact-diagonalization methods, we determined that $\chi=8$ is sufficient to represent the low-energy states of interest in this model. In this section, we describe the results for the spin-model in the absence of magnetic fields $\boldsymbol{B}\equiv(0,0,0)$. In the appendix, we provide additional results for the same system at finite magnetic fields.
	
	The QTN method starts with the CTN description of the eigenstates which are then compiled into quantum circuits, first using the unitary embeddings, and then compilation to a quantum computing gate-set as discussed in \ref{subsec:statepre}. For all variational compilations, we use a tolerance of $\varepsilon_C = 5\times 10^{-4}$ and a set of entangling gates that includes CNOT gates (with Y rotations) and generic SO4 gates. We then extract the average energy and compute the transition matrix elements as described in \ref{subsec:meas} for translationally invariant systems. The simulations of QTN circuits are done using the QASM Simulator in the Qiskit python package~\cite{Qiskit}. We stress here that we do not attempt to further optimize the QTN to larger bond dimensions beyond that obtained from the CTN since the focus of this work is to benchmark the algorithms developed for extracting observable properties by comparing them to CTN obtained values. In practice, we anticipate that further variational optimization of QTNs with larger bond dimension will be essential to generating higher quality representations of the eigenstates of the Hamiltonian. An in-depth analysis of the advantage of using QTN circuits will be a subject of subsequent studies.
	
	Let us now discuss the specifics for extracting the values of the different observables. The average energy is a sum of single site and two-site operators. For the static expectation values, we simulate the quantum circuits with $X$, $Y$ and $Z$ basis measurements for $n_{\rm shots}= 10^{4}$ shots each. The expected value of the Hamiltonian is then evaluated from individually extracted expectation values of single site operators  $\hat{S}^\alpha_j$ and two-site correlations $\hat{S}^\alpha_i\hat{S}^\alpha_j$. Notice that, because of the nature of terms present, in the sequential preparation scheme, we measure all the physical qubits in the same basis and the expected values of each of the operators can be extracted from the same measurement dataset. 
	 In \cref{fig:zeroB-spinhalf}~(a) we compare the extracted energies to the energies extracted from CTN methods (benchmark). We see that the variational approach well approximates the exact energies for ground and excited states.

	In \cref{fig:zeroB-spinhalf}~(b), we carefully analyze the error from measurement noise in the estimate of the energy values. While at smaller values of $n_{\rm shots}$ the relative error decreases by increasing the number of shots, the error-rate saturates at large $n_{\rm shots}$. This saturation error value is approximately the error set by the tolerance in the compilation of the unitary embeddings. This is because the fidelity of the compiled QTN, consisting of $L$ physical sites, is approximately given by $F\approx 1-L\varepsilon_C/(\chi d +1)$. In \cref{app:err bounds}, we provide details for this estimate as well as analytical bounds on the error in measured expected values of observables as it relates to the fidelity of the compilation of the unitary embeddings. In \cref{fig:zeroB-spinhalf}~(c) we show the quantitative separation in gate counts when compiling the unitaries to the native gate set. The variational approach for compilation pursued here provides a constant factor speedup (in gate counts) consistently for all of the low-lying excited states simulated here. Note that this constant factor is tunable, and is set by the choice of the tolerance $\varepsilon_C$ in the variational procedure. We remark here that the circuit depths obtained for the QTN representations for the ground and excited states are independent of the choice of state because the bond dimension is chosen to be the same for all the states. Generalizing to higher excited states would generically require higher bond dimensions and a corresponding  increase in gate counts. The scaling of the gate depth with the effective bond dimension is expected to be non-trivial and problem dependent; some initial analysis of the scaling for transverse field Ising models may be found in Ref.~\cite{WallPRA2022}. Furthermore, for implementation on noisy hardware, increasing the bond-dimension at the cost of added gate depth may lead to less accurate solutions due to gate errors. In order to identify the optimal choice for the bond-dimension and gate depths,  one must carefully examine this trade-off; we leave this investigation for future work.

	Having established that the variational unitary embeddings are sufficient to compute the energy spectra of the spin chain, we now test the algorithm developed for the transition elements in \cref{subsec:meas} for translationally invariant spin-half systems. As noted before, we can use this algorithm to compute the elements $\hat{O}_{ij;p}^{\alpha \alpha}$. In \cref{fig:zeroB-spinhalf}~(d), we compare the matrix element, $\hat{O}_{03;3}^{xx}$. Clearly, we see good agreement between the classical and quantum methods for computing the matrix elements.

	\subsection{Spin-3/2 Ring: Cr\texorpdfstring{$^{3+}_8$}{Cr^{3+}_8}  Molecule} \label{subsec:extensions}
	In this section let us discuss the application of our methods to a realistic spin Hamiltonian model, that describes the low-temperature physics of the Cr$^{3+}_8$ molecule~\cite{Baker2012}. We consider a ring of $L=8$ spin-3/2 particles fixing the following parameters in \cref{eq:Hamiltonian}: nearest neighbor coupling $J=1.46$~meV, single site anisotropy $D=-0.038$~meV, the Land\'{e} g-factor $g=1.98$ and the Bohr magneton $\mu_B=5.7883818060\times 10^{-2}$~meV/T. As noted in \cref{subsec:meas}, the Hilbert space of a spin 3/2 system can be encoded into two qubits, and thus, the QTN representation has two physical qubits. We set the bond dimension $\chi=32$ in the CTN calculation, which implies that the QTN representation must use $\log_2(32)=5$ bond qubits. In the following, we discuss our results for extracted energies and the magnetic response functions discussed in \cref{sec:model}. For numerical convenience, we use the exact QTN embeddings in this section.

	Measuring the energy requires the estimation of single site and two-site observables. This corresponds to extracting expected values from measurement statistics of two qubit and four qubit correlations respectively. We provide expressions for measuring $\braket{\hat{S}^x_j}$ and $\braket{\hat{S}^x_j\hat{S}^x_{j+1}}$,
	\begin{align}
		&\braket{\hat{S}^x_j}= \frac{\sqrt{3}}{2}\braket{(\hat{I}\hat{\sigma}^x)_j} + \frac{1}{2}\braket{(\hat{\sigma}^x\hat{\sigma}^x)_j} + \frac{1}{2}\braket{(\hat{\sigma}^y\hat{\sigma}^y)_j}\\
		&\braket{\hat{S}^x_j \hat{S}^x_{j+1}} =\nonumber\\
		& \frac{1}{4}\Big[ 3\braket{(\hat{I}\hat{\sigma}^x)_j(\hat{I}\hat{\sigma}^x)_{j+1}}+ \sqrt{3}\braket{(\hat{I}\hat{\sigma}^x)_j(\hat{\sigma}^x\hat{\sigma}^x)_{j+1}} \notag\\
		&\hphantom{\frac{1}{4}\Big[}+ \sqrt{3}\braket{(\hat{\sigma}^x\hat{\sigma}^x)_j(\hat{I}\hat{\sigma}^x)_{j+1}} + \braket{(\hat{\sigma}^x\hat{\sigma}^x)_j(\hat{\sigma}^x\hat{\sigma}^x)_{j+1}} \notag\\
		&\hphantom{\frac{1}{4}\Big[}+ \sqrt{3}\braket{(\hat{I}\hat{\sigma}^x)_j(\hat{\sigma}^y\hat{\sigma}^y)_{j+1}} + \sqrt{3}\braket{(\hat{\sigma}^y\hat{\sigma}^y)_j(\hat{I}\hat{\sigma}^x)_{j+1}} \notag\\
		&\hphantom{\frac{1}{4}\Big[}+\braket{(\hat{\sigma}^x\hat{\sigma}^x)_j(\hat{\sigma}^y\hat{\sigma}^y)_{j+1}} +\braket{(\hat{\sigma}^y\hat{\sigma}^y)_j(\hat{\sigma}^x\hat{\sigma}^x)_{j+1}}\nonumber\\
		&\hphantom{\frac{1}{4}\Big[}+ \braket{(\hat{\sigma}^y\hat{\sigma}^y)_j(\hat{\sigma}^y\hat{\sigma}^y)_{j+1}} \Big]
	\end{align}
	where $(\cdots)_j$ denotes the physical qubits representing the spin qudit at site $j$.
	
\begin{figure}[t]
		\centering
		\includegraphics[width=\linewidth]{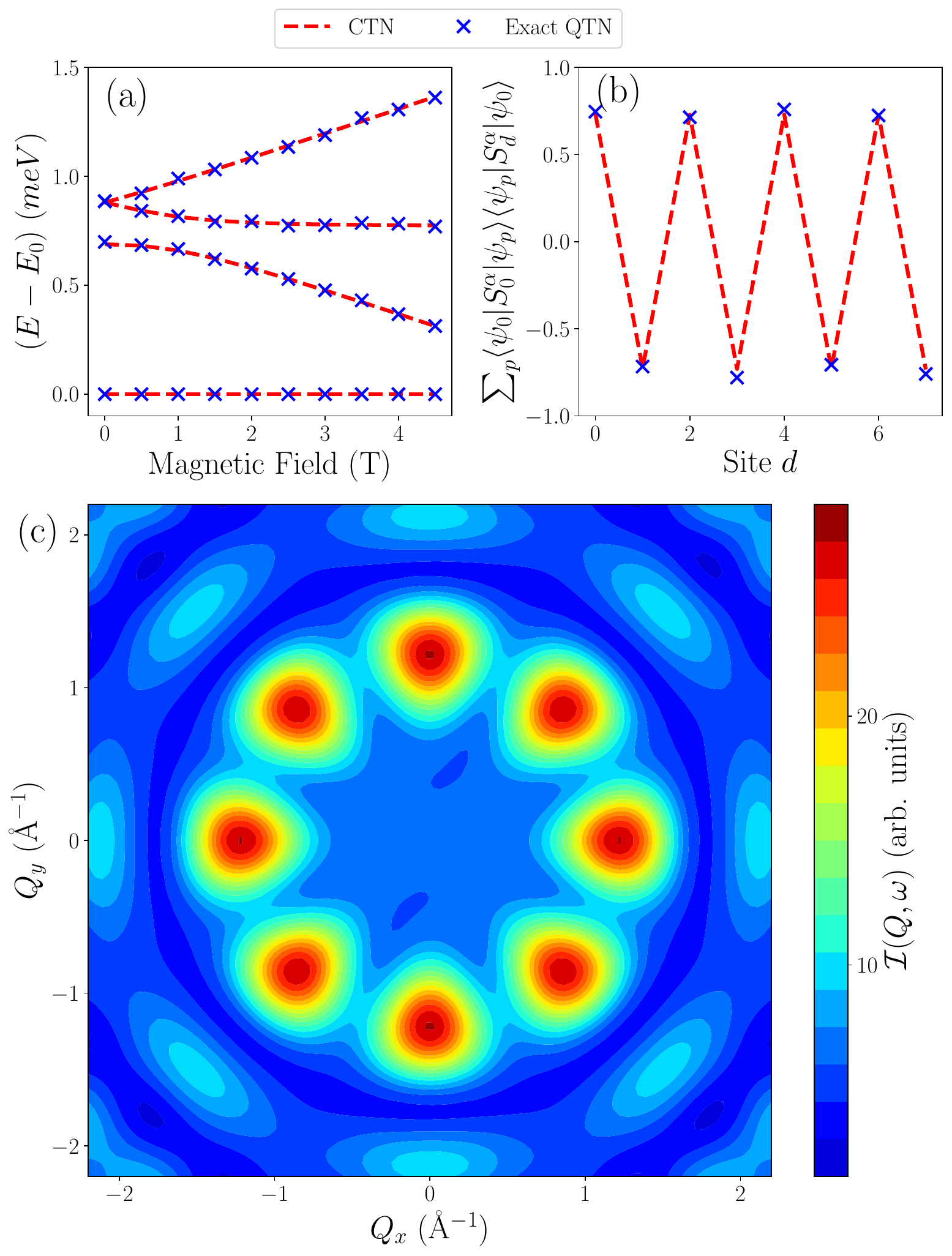}  
		\caption{\label{fig:spinTH} Benchmarking QTN algorithms against CTN evaluations. (a) Excitation energies of the Cr$_8$ molecule as a function of magnetic field. The ground state of the molecule is a spin-singlet state, and the first excited state manifold is the spin-triplet configuration. (b) Example of the magnetic dipole transition matrix elements between the ground state and the spin-triplet states, $\sum_{p=1}^3 O^{xx}_{0d;p}$. (c) Example of computation of the magnetic scattering intensity [See \cref{eq:SQdef}] using QTN methods.The calculated response function for the transition to the spin-triplet manifold with energy transfer $ \hbar\omega \in [0.5, 1] $ meV has good agreement with data from literature~\cite{Baker2012}.}
	\end{figure}

	The first four terms in the above expression can be obtained from the measurement outcomes of a circuit in which all physical qubits are readout in the $X$ basis, while the last requires only $Y$ measurements. However, the remaining terms involve measuring site $j$ in the $X$ basis and $j+1$ in the $Y$ basis, or vice versa, leading to a total of four distinct circuits that must be executed in order to reconstruct the expectation value. We illustrate this method by computing the ground and excited states as a function of magnetic field in ~\cref{fig:spinTH}~(a) using the exact QTN representation of the CTN states. For these results, we run the QTN circuits for $10^7$ shots each. As is clear from the figure, we have good agreement between the energies obtained from the CTN and QTN representations. The Cr$^{3+}_8$ molecule is characterized by a spin-singlet ground state ($S=0$), and a first-excited state manifold that is the spin-triplet configuration ($S=1$). The effect of the magnetic field is described by the Zeeman term, which causes the states of the first excited state manifold to split in energy.
	
	Next, we discuss computing the magnetic dipole transition matrix elements [e.g. \cref{fig:spinTH}~(b)] as well as the response function for unpolarized neutron scattering [e.g. \cref{fig:spinTH}~(c)]. We use the generalized swap-test approach to computing the necessary matrix elements for extracting the magnetic scattering intensity defined in \cref{eq:SQdef}. The QTN swap tests are run with $n_{\rm shots}=5000$ per circuit, which is sufficient to get accurate estimates of the overlap. We compute the magnetic response function at zero magnetic field for transitions between the ground state and spin-triplet states, ($p=1,2,3$), which can be probed using neutron scattering experiments (See Ref \onlinecite{Baker2012}). Note that the scattering intensity is obtained by averaging the intensity from two molecular orientations. We provide details for the form factors and orientations of the Cr$^{3+}$ ions in \cref{app:spin3/2}. The extracted response function has good agreement with experimentally observed data from Ref. \onlinecite{Baker2012}. 

\section{Discussion and Outlook}\label{sec:discussion}
	We have presented a tensor network based approach to evaluating dynamic correlation functions on a quantum computer. We have shown that classical tensor network data structures representing ground and excited states of a quantum spin model can be transformed into a quantum circuit that can be evaluated on a quantum computer. To summarize the approach, the quantum state is prepared sequentially (site by site) using a set of qubits for the physical dimension ($\log_2 d$ qubits) representing the physical spins of the system, and a set of bond qubits ($\log_2\chi$ qubits) that represent the ancillary bond dimension which characterize the amount of entanglement in the state.

	We have outlined different approaches to extracting static and dynamic correlation functions. In particular, we have developed two novel approaches to compute particular magnetic dipole transition matrix elements. The first approach exploits the Fourier transformation to convert the matrix elements into an overlap probability measurement, and can be thought of as a quantum circuit version of a matrix product operator.  The scheme was worked out in detail for spin-half systems. The second approach, based on a generalization of the SWAP test, allows one to extract this matrix elements for arbitrary quantum states encoded in the sequential preparation scheme. We have shown the general applicability of both approaches by considering exemplar spin-1/2 and spin-3/2 Heisenberg models. The latter system is a realistic model of a magnetic nano-molecule, and our results compare favorably with experimental measurements of this system.

	We have developed an approach that allows using near-term quantum hardware with measurement and active reset capabilities to simulate static and dynamic properties of materials. In particular, trapped-ion hardware has already been demonstrated using the sequential preparation scheme for ground-state energy calculations as well as machine learning applications~\cite{foss2021entanglement,WallPRA2022}. These systems can now readily be used to implement the algorithms discussed in this work to extract dynamic correlation functions.

	There are several immediately relevant extensions that can be built up from the results discussed in this paper. The variational approach discussed here relies on the exact CTN representation of the MPS state but demonstrates that QTNs with modest gate depth can approximate these optimal CTNs with errors comparable to that expected from current, noisy quantum computers. This approach is readily generalizable to variationally optimize QTNs with larger bond dimensions by using the CTN as a pre-conditioner for the ground as well as excited states. It will be essential to also study the effect of gate errors on the accuracy of the measured observables, and the robustness of the variational compilation strategy to these errors. Other interesting directions to explore are fermionic models where various response functions may be related to quantum transport properties. Finally, while we have focused on MPS states, one may consider generalizations to other tensor network topologies to simulate novel many-body quantum states such as quantum critical states.

\section{Acknowledgements}
M.L.W., A.R., J.S.V and P.T. would like to acknowledge funding from the Internal Research and Development program of the Johns Hopkins University Applied Physics Laboratory. The authors acknowledge valuable discussions with Tim Reeder. C.B. was  supported by the U.S. DOE Office of Science, Office of Basic Energy Sciences, Materials Sciences and Engineering Division DE-SC0024469 and by the Gordon and Betty Moore foundation through GBMF9456.

\appendix

\section{Error analysis: Fidelity and Observables}
\label{app:err bounds}
In this appendix, we provide an estimate of the state fidelity $F$ between the variational compilation of the  QTN circuits and the exact representation which is then used to bound the error in any measured observable. 

We start by relating the fidelity between the two states to the tolerance $\varepsilon_C$ chosen for the variational compiler. We represent the set of unitary operators for the exact (non-variational) state preparation by $\{\hat U_{\rm L}^{[j]}\}$, and the compiled unitary operators $\{\hat U_{\rm L,c}^{[j]}\}$ with $j=1\cdots L$; the dimensions of these operators are at most $n=d\chi$. We also represent the corresponding full state preparation unitary (from successive applications of the $L$ unitaries) as $U$ and $U_{\rm c}$,
\begin{align}
	\ket{\psi}=\hat{U}\ket{0},\ \ket{\psi_{\rm c}}=\hat{U}_{\rm c}\ket{0}
\end{align}
where $\ket{0}$  denotes all physical and bond qubits. The tolerance of the variational compiler sets the  2-norm distance between the unitary operators for each site which then bounds the distance between the full unitary operators,
\begin{align}
	\left\|\hat U_{\rm L,c}^{[j]}-\hat U_{\rm L}^{[j]}\right\|_2^2&\leq \varepsilon_C,\\
	\Rightarrow \left\|\hat U_{\rm c}-\hat U\right\|_2^2 &\leq L\varepsilon_C.
\end{align}
where $\|\cdots\|_p$ denotes the Schatten $p$-norm. This distance metric can be used to derive a lower bound to the entanglement fidelity between the two operators. We have,
\begin{align}
	\Rightarrow \Tr\left[\left(U_{\rm c}-\hat U\right)^\dagger\left(U_{\rm c}-\hat U\right)\right]&\leq L\varepsilon_C,\\
	\Rightarrow \Re\left[\Tr\left[U_{\rm c}^\dagger\hat U\right]\right]&\geq n-\frac{1}{2}L\varepsilon_C,
\end{align}
where $\Re[\cdots]$ denotes the real part, and we used $\Tr[\hat I]=n$. Using this, we can lower bound the entanglement fidelity of these two unitary operators which is given by,
\begin{align}
	F_{\rm e}=\frac{1}{n^2} \left|\Tr\left[\hat U^\dagger_{\rm c}\hat U\right]\right|^2\geq 1-L\frac{\varepsilon_C}{n},
\end{align}
where we use $|A|\geq \Re[A]$. Now, The entanglement fidelity is related to the average fidelity $\overline{F}$~\cite{Nielsen2002}, which is given by, 
\begin{align}
	\overline{F}=\frac{nF_e+1}{n+1}=1-\frac{L}{n+1}\varepsilon_C.
\end{align}
We use this average fidelity as an approximate estimate of the true fidelity between the two states. Thus we arrive at the estimate,
\begin{align}
	F=\left|\braket{\psi_{\rm c}|\psi}\right|^2 \sim 1-\frac{L}{n+1}\varepsilon_C.
\end{align}
The fidelity can be related to the trace distance between the two states $\delta=\frac{1}{2}\|\ket{\psi}\bra{\psi}-\ket{\psi_{\rm c}}\bra{\psi_{\rm c}}\|_1 = \sqrt{1-F}$.

Let us now calculate a bound on the error in calculating the expected value of an arbitrary Hermitian observable $\hat{O}$ using the state $\ket{\psi_{\rm c}}$.  The absolute error in the expected value is given by, 
\begin{align}
	\varepsilon^{\rm abs}_O&=\left|\braket{\psi|\hat{O}|\psi} - \braket{\psi_{\rm c}|\hat{O}|\psi_{\rm c}}\right| \nonumber\\
	&= \left|\Tr\left[\hat{O}\left(\ket{\psi}\bra{\psi} - \ket{\psi_{\rm c}}\bra{\psi_{\rm c}}\right)\right]\right| \label{eq:abserrordef}
\end{align}
Using the H\"{o}lder's inequality for matrices, we can upper bound this error [R.H.S in \cref{eq:abserrordef}] in the expected value,
\begin{align}
	\left|\Tr\left[\hat{O}\left(\ket{\psi}\bra{\psi} - \ket{\psi_{\rm c}}\bra{\psi_{\rm c}}\right)\right]\right|&\leq\|\hat{O}\|_\infty\left\|\ket{\psi}\bra{\psi} - \ket{\psi_{\rm c}}\bra{\psi_{\rm c}}\right\|_1.
	\label{eq:obs bound}\\
	\Rightarrow \varepsilon^{\rm abs}_O&\leq 2\|\hat{O}\|_{\infty}\sqrt{1 - F}
\end{align}
%  Now, using the definition of the trace distance between the two states and its relationship to the fidelity, we can simplify the expression for the absolute error,
% \begin{equation}
% 	\varepsilon_O
% \end{equation}
%As an example consider the observable $S_{\alpha}^iS_{\alpha}^{i+1}$, $\alpha\in{x,y,z}$, for spin-1/2 operators. In this case,$\|S_{\alpha}^iS_{\alpha}^{i+1}\|_\infty=\frac{1}{4}$, which implies,
We estimate the relative error as $\varepsilon_O\sim\varepsilon^{\rm abs}_O/\|O\|_{\infty}$,
\begin{equation}
	\varepsilon_O \leq 2\sqrt{1 - F} \sim 2\sqrt{\frac{L\varepsilon_C}{(d\chi+1)}} 
\end{equation}
In our numerical simulations of the spin-1/2 chain in \cref{subsec:spin 1/2} these parameters are: $\varepsilon_C=5\times 10^{-4}$, $L=8$, $\chi=8$ and $d=2$ which corresponds to a relative error of approximately $\varepsilon_O\approx 0.03$ consistent with the errors observed in simulation.
\begin{figure}
	\begin{center}
		\includegraphics[width=0.95\columnwidth]{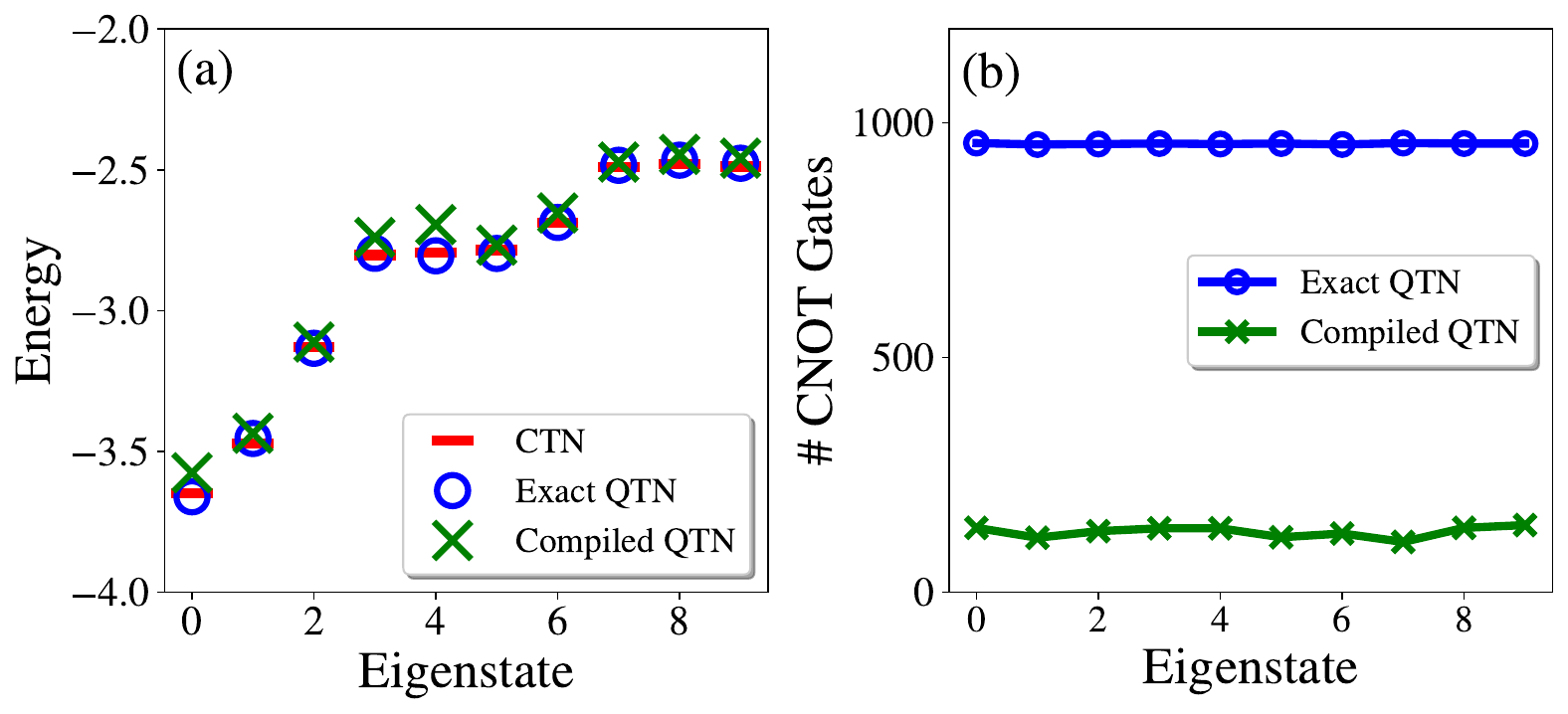}  
		\caption{\label{fig:nonzeroB-spinhalf} Comparing CTN and QTN methods for evaluating observables in the spin-half Heisenberg chain with $\boldsymbol{B} = 3 \cdot (0.653, 0 , 0.758)$ T. (a)  First 10 Energy levels (in meV) with $n_{\mathrm{shots}}=10^{4}$. (b) Variational compilation of the QTN circuits. Gate counts are substantially reduced for variational compilation methods in comparison with exact (non-variational) methods. }
	\end{center}
\end{figure}

\section{Additional Examples: Spin-half Heisenberg chain in finite magnetic fields.}\label{app:finite-mag-spin-1/2}
In this appendix, we provide an additional example of the spin-half Heisenberg chain in a non-zero magnetic field. The magnetic field is set to $\boldsymbol{B} = 3 \cdot (0.653, 0 , 0.758)$ T , with $g=1.98$. In \cref{fig:nonzeroB-spinhalf} (a) we provide a comparison of the extracted energies from the QTN representations of the low-lying eigenstates of the Hamiltonian. In \cref{fig:nonzeroB-spinhalf} (b) we show the gate count comparison between exact and variational compilation methods. Again, we see a constant factor improvement in the gate counts for representing the QTN representations.

\section{Additional details for \texorpdfstring{Cr$_8$}{Cr8}}\label{app:spin3/2}
\begin{figure}[h]
	\includegraphics[width=\linewidth]{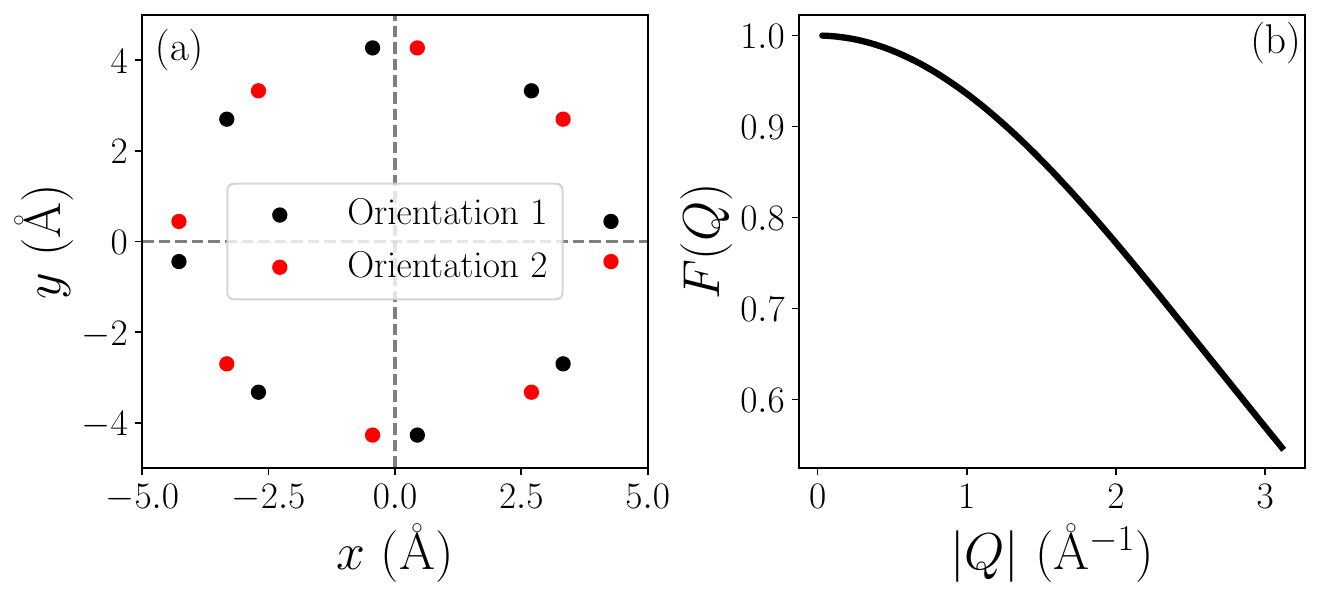}
	\caption{\label{fig:formfacs}(a) Positions of the individual Cr$^{3+}$ ions in the z-plane taken to compute the dot product with momentum vectors, $(Q_x,Q_y)$ in the expressions for the magnetic response function. (b) Form factors evaluated from the expression taken from Ref.~\onlinecite{formfacs}}
\end{figure}

In this appendix, we provide some additional details for computation of inelastic neutron scattering cross-section from \cref{sec:model}. We consider the ions in the ring-shaped molecule to be in the x-y plane with the coordinates shown in \cref{fig:formfacs}~(a). There are two orientations, the scattering intensity is averaged over both of these orientations.

The expressions for the form factor are taken from Ref.~\onlinecite{formfacs} and is given by,
\begin{align}
	\label{eq:formfac0} F(s)=\langle j_0(s)\rangle+\frac{2-g}{g}\langle j_2(s)\rangle
\end{align}
where, $s=|\boldsymbol{Q}|/4\pi$ with $\boldsymbol{Q}$ being the momentum wavevector. These functions are expanded as a series of exponential functions,
\begin{align}
\label{eq:formfac1}\langle j_0(s)\rangle&=Ae^{-a s^2}+Be^{-b s^2}+Ce^{-c s^2}+D,\\
\label{eq:formfac2}\langle j_2(s)\rangle&=s^2\left(Ae^{-a s^2}+Be^{-b s^2}+Ce^{-c s^2}+D\right),
\end{align}
where the coefficients for the specific ion (Cr$^{3+}$ (a 3d transition element)) are provided in \cref{tab:formfacs}.
\begin{table}[h]
	\begin{center}
		\begin{tabular}{|c|c|c|c|c|c|c|c|c|}
			\hline
			\hphantom{a} & A & a & B & b & C & c & D & e \\
			\hline
		$j_0$& -0.3094 & 0.0274 & 0.3680 & 17.0355 & 0.6559 & 6.5236 & 0.2856 & 0.0436 \\
		$j_2$& 1.6262&15.0656 & 2.0618 & 6.2842 &0.5281 &2.3680 &0.0023&0.0263\\
		\hline
		\end{tabular}
	\end{center}
	\caption{Table of constants for the form factor calculation in \cref{eq:formfac0,eq:formfac1,eq:formfac2}.\label{tab:formfacs}}
\end{table}

	\bibliography{QTN_refs} 
\end{document}